\documentstyle[12pt]{article}
\def\hybrid{\topmargin -20pt  \oddsidemargin 0pt
      \headheight 0pt   \headsep 0pt
      \textwidth 6.25in 
      \textheight 9.5in 
      \marginparwidth .875in
      \parskip 5pt plus 1pt   \jot = 1.5ex}

\hybrid

\begin{document}
\def\x{\times}
\def\beq{\begin{equation}}
\def\eeq{\end{equation}}
\def\beqa{\begin{eqnarray}}
\def\eeqa{\end{eqnarray}}
\def\D{ {\cal D}}
\def\L{ {\cal L}}
\def\C{ {\cal C}}
\def\N{ {\cal N}}
\def\calE{{\cal E}}
\def\lin{{\rm lin}}
\def\Tr{{\rm Tr}}
\def\mxth{\mathsurround=0pt }
\def\xversim#1#2{\lower2.pt\vbox{\baselineskip0pt \lineskip-.5pt
x  \ialign{$\mxth#1\hfil##\hfil$\crcr#2\crcr\sim\crcr}}}
\def\simgr{\mathrel{\mathpalette\xversim >}}
\def\simle{\mathrel{\mathpalette\xversim <}}

\def\a{\alpha}
\def\b{\beta}
\def\dota{ {\dot{\alpha}} }
\def\lag{Lagrangian}
\def\Kahler{K\"{a}hler}
\def\kahler{K\"{a}hler}
\def\A{ {\cal A}}
\def\C{ {\cal C}}
\def\D{ {\cal D}}
\def\F{{\cal F}}
\def\L{ {\cal L}}

\def\R{ {\cal R}}
\def\x{ \times }
\def\beq{\begin{equation}}
\def\eeq{\end{equation}}
\def\beqa{\begin{eqnarray}}
\def\eeqa{\end{eqnarray}}

\sloppy
\newcommand{\be}{\begin{equation}}
\newcommand{\eq}{\end{equation}}
\newcommand{\ov}{\overline}
\newcommand{\un}{\underline}
\newcommand{\p}{\partial}
\newcommand{\la}{\langle}
\newcommand{\ra}{\rangle}
\newcommand{\bl}{\boldmath}
\newcommand{\ds}{\displaystyle}
\newcommand{\nl}{\newline}
\newcommand{\Nzahl}{{\bf N}  }
\newcommand{\zzahl}{ {\bf Z} }
\newcommand{\Zzahl}{ {\bf Z} }
\newcommand{\Qzahl}{ {\bf Q}  }
\newcommand{\Rzahl}{ {\bf R} }
\newcommand{\Czahl}{ {\bf C} }
\newcommand{\wt}{\widetilde}
\newcommand{\wh}{\widehat}
\newcommand{\fs}[1]{\mbox{\scriptsize \bf #1}}
\newcommand{\ft}[1]{\mbox{\tiny \bf #1}}
\newtheorem{satz}{Satz}[section]
\newenvironment{Satz}{\begin{satz} \sf}{\end{satz}}
\newtheorem{definition}{Definition}[section]
\newenvironment{Definition}{\begin{definition} \rm}{\end{definition}}
\newtheorem{bem}{Bemerkung}
\newenvironment{Bem}{\begin{bem} \rm}{\end{bem}}
\newtheorem{bsp}{Beispiel}
\newenvironment{Bsp}{\begin{bsp} \rm}{\end{bsp}}
\renewcommand{\arraystretch}{1.5}



\renewcommand{\thesection}{\arabic{section}}
\renewcommand{\theequation}{\thesection.\arabic{equation}}

\parindent0em

\def\S4{\frac{SO(4,2)}{SO(4) \otimes SO(2)}}
\def\P3{\frac{SO(3,2)}{SO(3) \otimes SO(2)}}
\def\MGd{\frac{SO(r,p)}{SO(r) \otimes SO(p)}}
\def\SOd{\frac{SO(r,2)}{SO(r) \otimes SO(2)}}
\def\SO2{\frac{SO(2,2)}{SO(2) \otimes SO(2)}}
\def\SUm{\frac{SU(n,m)}{SU(n) \otimes SU(m) \otimes U(1)}}
\def\SUS{\frac{SU(n,1)}{SU(n) \otimes U(1)}}
\def\SK{\frac{SU(2,1)}{SU(2) \otimes U(1)}}
\def\SU{\frac{ SU(1,1)}{U(1)}}

\begin{titlepage}
\begin{center}

\hfill HUB-EP-95/33\\
\hfill CERN-TH/95-341\\
\hfill SNUTP-95/095\\
\hfill hep-th/9512129\\

\vskip .1in

{\bf BPS Spectra and
Non-Perturbative Gravitational Couplings in $N=2,4$
Supersymmetric String Theories}

\vskip .2in

{\bf Gabriel Lopes Cardoso$^a$, Gottfried Curio$^b$,
Dieter L\"ust$^b$, Thomas Mohaupt$^b$ \\
and \\
Soo-Jong Rey$^c$}\footnote{email: cardoso@surya15.cern.ch,
	curio@qft2.physik.hu-berlin.de,
	luest@qft1.physik.hu-berlin.de,\\
	mohaupt@qft2.physik.hu-berlin.de, sjrey@phyb.snu.ac.kr}
\\
\vskip 1.2cm
$^a${\em Theory Division, CERN, CH-1211 Geneva 23, Switzerland}

$^b${\em Humboldt-Universit\"at zu Berlin,
Institut f\"ur Physik\\
D-10115 Berlin, Germany}

$^c${\em Department of Physics, Seoul National University,
Seoul 151-742 Korea}

\vskip .1in

\end{center}

\vskip .2in

\begin{center} {\bf ABSTRACT } \end{center}
\begin{quotation}\noindent
We study the BPS spectrum in $D=4, N=4$ heterotic string compactifications,
with some emphasis on
intermediate $N=4$ BPS states.  These intermediate states,
which can become short in $N=2$ compactifications, are  crucial
for establishing an $S-T$ exchange symmetry in $N=2$ compactifications.
We discuss the implications of a possible $S-T$ exchange symmetry for
the $N=2$ BPS spectrum. Then
we present the exact result for the 1-loop corrections to
gravitational couplings in one of the heterotic $N=2$ models
recently discussed by Harvey and Moore.
We conjecture this model to have an $S-T$ exchange symmetry.  This
exchange symmetry
can then be used to evaluate non-perturbative corrections to
gravitational couplings in some of the non-perturbative regions (chambers)
in this particular model and also in other heterotic models.
\end{quotation}
December 1995\\
\end{titlepage}
\vfill
\eject

\newpage

\section{Introduction}

Recently, some major progress has been obtained in the understanding
of non-perturbative dynamics in field theories and string theories
with extended supersymmetry
\cite{SW1,SW2,KLTY,KLT,HullTown,Wit,KacVaf,FerHarvStrVa,KLM,CLM,AntoPartou,KKLMV}.
One important feature of these theories
is the existence of BPS states. These BPS states play an important
role in understanding duality symmetries
and non-perturbative effects in string theory in various dimensions.
They are, for instance, essential to the resolution of the
conifold singularity in type II string theory \cite{Strom}.
BPS states also play a central role in 1-loop
threshold corrections
to gauge and gravitational couplings in $N=2$ heterotic string
compactifications, as shown recently in \cite{HarvMoo}.

In the context of $D=4,N=4$ compactifications,
BPS states also play a crucial role in
tests \cite{Sen1} of the conjectured strong/weak coupling
$SL(2,{\bf Z})_S$ duality
\cite{sduality,SchSen,Sen2}
in toroidal compactifications of the heterotic string.
Moreover, the
conjectured
string/string/string triality \cite{DuffLiuRahm}
interchanges the BPS spectrum of the heterotic
theory with the BPS spectrum of the type II theory.
In an $N=4$ theory, BPS states can
either fall into short or into intermediate
multiplets.
 In going from the heterotic to the type IIA side, for
example, the four dimensional axion/dilaton field $S$ gets
interchanged with the complex K\"{a}hler modulus $T$ of the 2-torus
on which the type IIA theory has been compactified on \cite{duff1,duff2}.  
Thus, it is
under the exchange of $S$ and $T$ that the BPS spectrum of
the heterotic and the type IIA string gets mapped into each other.
The BPS mass spectrum of the heterotic(type IIA) string is, however,
not symmetric under this exchange of $S$ and $T$.  This is due to
the fact that BPS masses in $D=4, N=4$ compactifications are
given by the maximum of the 2 central charges $|Z_1|^2$ and
$|Z_2|^2$ of the $N=4$ supersymmetry algebra \cite{FerSaZu}.

On the other hand,
states, which from the $N=4$ point of view are intermediate,
are
actually short from the $N=2$ point of view.  This then leads
to the possibility that the BPS spectrum of certain
$N=2$ heterotic compactifications is actually symmetric under
the exchange of $S$ and $T$.
If such symmetry exists a lot of information
about the BPS spectrum at strong coupling can be obtained,
in particular about those BPS states which can become
massless at specific points in the
moduli space. Assuming that the contributions
to the associated gravitational couplings are due to BPS
states only (as was shown to be the case at 1-loop for some
classes of compactifications
in \cite{HarvMoo}),
it follows that these
gravitational couplings
should also exhibit such an $S \leftrightarrow T$ exchange symmetry.
The evaluation of non-perturbative
corrections to gravitational couplings is, however, very difficult.
The existence of an exchange symmetry
$S \leftrightarrow T$ is extremly helpful in that it allows
for the evaluation of non-perturbative corrections to gravitational couplings
in some of the non-perturbative regions (chambers) in moduli space.
This is achieved by taking the known result for the 1-loop correction
in some perturbative region (chamber) of moduli space and applying
the exchange symmetry to it.  Three
examples will be discussed in
this paper, namely the 2 parameter model $P_{1,1,2,2,6}(12)$ of
\cite{CanOsFoKaMo},
the 3 parameter model $P_{1,1,2,8,12}(24)$ \cite{KacVaf,KKLMV} (for these two
models an exchange symmetry $S\leftrightarrow T$ has been observed in
\cite{KLM})
and the $s=0$ model of \cite{HarvMoo}
(for this example we conjecture that there too is such an exchange
symmetry).

The paper is organised as follows.
In section 2 we introduce orbits for short and intermediate multiplets
in $D=4,N=4$ heterotic string compactifications and we show how
they get mapped into each other under string/string/string
triality.
In section 3 we discuss BPS states in the context of
$D=4,N=2$ heterotic string compactifications and show that
states, which from the $N=4$ point of view are intermediate,
actually play an important role in the correct evaluation of
non-perturbative effects such as non-perturbative monodromies.
We also discuss exchange symmetries of the
type $S \leftrightarrow T$
in the 2 and 3 parameter models $P_{1,1,2,2,6}(12)$ and
$P_{1,1,2,8,12}(24)$.
In section 4 we introduce an $N=4$ free energy as a sum over
$N=4$ BPS states and suggest that it should be identified
with the partition function of topologically twisted $N=4$
string compactifications.  In section 5 we introduce an
$N=2$ free energy as a sum over $N=2$ BPS states and argue
that it should be identified with the heterotic
holomorphic gravitational function ${\cal F}_{grav}$.
We discuss 1-loop corrections to the gravitational coupling
and compute them exactly in the $s=0$
model of \cite{HarvMoo}.  We then argue that this model
possesses an $S \leftrightarrow T$ exchange symmetry and
use it to compute non-perturbative corrections to the
gravitational coupling in some non-perturbative regions of moduli
space.  We also discuss the
2 parameter model $P_{1,1,2,2,6}(12)$ of
\cite{CanOsFoKaMo} and compute the associated holomorphic
gravitational coupling
in the decompactification limit $T \rightarrow \infty$.
Finally, appendices A and B contain a more detailed discussion
of some of the issues discussed in section 2.

\section{The $N=4$ BPS spectrum}

\subsection{The truncation of the mass formula}

In this section we recall the  BPS mass formulae for four-dimensional
string
theories with $N=4$ space-time supersymmetry
\cite{SchSen,DuffLiuRahm}. Specifically, we first consider
the heterotic string compactified on a six-dimensional torus.
In $N=4$ supersymmetry, there are in general
two central charges $Z_1$ and $Z_2$.
There exist two kinds of massive BPS
multiplets, namely first the short multiplets which
saturate two BPS bounds (the associated soliton background solutions
preserve $1/2$ of the supersymmetries in $N=4$),
i.e.
\beqa m_S^2=|Z_1|^2=|Z_2|^2;
\label{short}
\eeqa
the short vector multiplets contain maximal spin one. Second there are
the intermediate multiplets which
saturate only one BPS bound
and contain
maximal spin $3/2$
(the associated solitonic backgrounds
preserve only one supersymmetry in $N=4$), i.e.
\beqa m_I^2={\rm Max}(|Z_1|^2,|Z_2|^2).
\label{interm}
\eeqa
The BPS masses are functions of the moduli parameter as well as functions
of the dilaton-axion field
$S=\frac{4 \pi}{g^2} - i \frac{\theta}{2\pi}=e^{-\phi}-ia$. Specifically,
the two central charges $Z_{1,2}$ have the following form
\cite{Cvetic,DuffLiuRahm}
\beqa
|Z_{1,2}|^2=\vec Q^2+\vec P^2\pm 2
\sqrt{\vec Q^2\vec P^2-(\vec Q\cdot\vec P
)^2},\label{duffcve}
\eeqa
where $\vec Q$ and $\vec P$
are the (6-dimensional) electric and magnetic charge
vectors which depend on the moduli and on $\phi,a$. One sees
that for short vector multiplets, for with $|Z_1|=|Z_2|$, the square root
term in (\ref{duffcve}) must be absent,
which is satisfied for parallel electric and
magnetic charge vectors. In this case the BPS masses agree with the formula
of Schwarz and Sen \cite{SchSen}.

In a general compactification on a six-dimensional torus $T^6$ the
moduli fields locally parametrize a homogeneous coset space
$SO(6,22)/(SO(6)\times SO(22))$.
In terms of these moduli fields, the
two central charges are then given\footnote{We are using
the notation of \cite{Sen2,DuffLiuRahm}.} by \cite{DuffLiuRahm}
\beqa
|Z_{1,2}|^2 = \frac{1}{16} \left( \gamma^T {\cal M}
( M + L ) \gamma \pm
\sqrt{(\gamma^T \epsilon \gamma)_{ab}
(\gamma^T \epsilon \gamma)_{cd} ( M+L)_{ac} (M+L)_{bd}} \right)
\label{centralchar}
\eeqa
where $\gamma^T=(\alpha,\beta)$.
Let us from now on restrict the discussion by considering only
an $SO(2,2)$ subspace which corresponds to two complex moduli fields $T$
and $U$. This means that we will only consider the moduli degrees of freedom
of a two-dimensional two-torus $T_2$. ($\vec Q$ and $\vec P$ are now
two-dimensional vectors.)
Then,
converting to a basis
where $L$ has diagonal form, $\check{L} = T^{-1} L T,
\check{M} = T^{-1} M T, \check{M} + \check{L} =
2 \phi \phi^T = \varphi \varphi^{\dag} + {\bar \varphi} \varphi^T$,
the two central charges can be written as
\beqa
|Z_{1,2}|^2 = \frac{1}{16} \left( \check{\gamma}^T {\cal M}
( \varphi \varphi^{\dagger} +
\bar{\varphi} \varphi^T ) \check{\gamma} \pm 2
\sqrt{(\check{\gamma}^T \epsilon \check{\gamma})_{ab}
(\check{\gamma}^T \epsilon \check{\gamma})_{cd}
 {\cal R}_{ac}
{\cal R}_{bd}} \right)
\eeqa
where
$\check{\gamma}^T = (\check{\alpha}, \check{\beta})=
(T^{-1} \alpha, T^{-1} \beta)$ and
where $ {\cal R}_{ac} = \frac{1}{2}
(\varphi \varphi^{\dagger} + {\bar \varphi}
\varphi^T)_{ac}$.
Using that $(\check{\gamma}^T \epsilon \check{\gamma})_{ab}
= \check{\alpha}_a \check{\beta}_b - \check{\alpha}_b \check{\beta}_a$
it follows that
\beqa
|Z_{1,2}|^2 &=& \frac{1}{16} \left( \check{\gamma}^T {\cal M}
( \varphi \varphi^{\dagger} +
\bar{\varphi} \varphi^T ) \check{\gamma} \pm 4
i \check{\alpha}^T {\cal I} \check{\beta} \right) \nonumber\\
&=& \frac{1}{4(S + \bar{S})} \left( \check{\alpha}^T {\cal R}
\check{\alpha}
+ S \bar{S} \check{\beta}^T {\cal R} \check{\beta} +
i(S-\bar{S}) \check{\alpha}^T {\cal R}
\check{\beta} \right. \nonumber\\
& \pm& i(S+\bar{S})
\check{\alpha}^T {\cal I} \check{\beta} \left.  \right)
\eeqa
where $ {\cal I} = \frac{1}{2} (
\varphi \varphi^{\dagger} - {\bar \varphi} \varphi^T)$.
The central charges $|Z_{1,2}|^2$ can
finally also be rewritten into
\beqa
|Z_{1,2}|^2
&=& \frac{1}{4(S + \bar{S})} \left( \check{\alpha}^T {\cal R}
\check{\alpha}
+ S \bar{S} \check{\beta}^T {\cal R} \check{\beta}
\pm i(S-\bar{S}) \check{\alpha}^T {\cal R}
\check{\beta} \pm i(S+\bar{S}) \check{\alpha}^T {\cal I}
\check{\beta}   \right) \nonumber\\
&=& \frac{1}{4(S + {\bar S})(T + {\bar T})(U + {\bar U})
}|{\cal M}_{1,2}|^2 \nonumber\\
{\cal M}_1 &=& \left( \check{M}_I + i S \check{N}_I \right)
\check{P}^I \nonumber\\
{\cal M}_2 &=& \left( \check{M}_I - i {\bar S} \check{N}_I \right)
\check{P}^I
\eeqa
where
\beqa
\check{P}^0 = T + U  \;\;,\;\; \check{P}^1 = i (1+TU) \nonumber\\
\check{P}^2 = T - U  \;\;,\;\; \check{P}^3 = -i (1-TU)
\eeqa
and where $\check{M}=\check{\alpha}, \check{N}=\check{\beta}$.
Here, the $\check{M}_I$ ($I=0,\dots ,3$)
are the integer electric charge quantum
numbers of the Abelian gauge group $U(1)^4$ and
the $\check{N}_I$ are the corresponding
integer magnetic quantum numbers.

Note that $|Z_2|^2$ can be obtained from $|Z_1|^2$
by $S \leftrightarrow \bar{S}, \check{N}_I \rightarrow -
\check{N}_I$.  This amounts to complex conjugating $
\check{M}_I + i S \check{N}_I $.

Finally, rotating the $\check{P}^I$ into $\hat{P}=(1, -TU, iT,iU)^T$
\beqa
\check{P} = A \hat{P}  \;\;,\;\;
A = i \left( \begin{array}{cccc}
0&0&-1&-1\\
1&-1&0&0 \\
0&0&-1&1\\
-1&-1&0&0\\
\end{array} \right)
\eeqa
gives that
\beqa
|Z_{1,2}|^2
&=& \frac{1}{4(S + {\bar S})(T + {\bar T})(U + {\bar U})
}|{\cal M}_{1,2}|^2 \nonumber\\
{\cal M}_1 &=& \left( \hat{M}_I + i S \hat{N}_I \right)
\hat{P}^I \nonumber\\
{\cal M}_2 &=& \left( \hat{M}_I - i {\bar S} \hat{N}_I \right) \hat{P}^I
\label{massphy}
\eeqa
where $\hat{M} = A^T \check{M}, \hat{N} = A^T \check{N}$.

Note that
\beqa
\Delta Z^2&=& |Z_1|^2-|Z_2|^2=
4\sqrt{\vec
Q^2\vec P^2-(\vec Q\cdot\vec P)^2}=
i {(\hat N_J \hat{P}^J)( \hat M_I  \bar{\hat{P}}^I)\over 4(T+\bar T)(U+\bar U)}
-{\rm h.c.}
\nonumber\\
&=&i {(\hat N_0-\hat N_1TU+i\hat N_2T+i\hat N_3U)(\hat M_0-\hat M_1
\bar T\bar U-i\hat M_2\bar T-i\hat M_3\bar U)\over 4(T+\bar T)(U+\bar U)}
-{\rm h.c.} \nonumber\\
\label{deltaz}
\eeqa
is independent of $S$ and only depends on the moduli $T$ and $U$.

The BPS mass formula
(\ref{massphy}) is invariant  under the perturbative $T$-duality group
$SL(2,{\bf Z})_T\times SL(2,{\bf Z})_U
\times{\bf Z}_2^{T\leftrightarrow U}$; for example $SL(2,{\bf Z})_T$,
$T\rightarrow {aT-ib\over icT+d}$, acts on the electric and magnetic charges
as
\beqa
\pmatrix{\hat M_2\cr \hat M_0\cr}\rightarrow
\pmatrix{a&c\cr b&d\cr}\pmatrix{\hat M_2\cr
\hat M_0\cr},\label{tdual}
\eeqa
where the vectors $\pmatrix{\hat M_1\cr \hat M_3\cr}$,
$\pmatrix{\hat N_2\cr \hat N_0}$ and
$\pmatrix{\hat N_1\cr \hat N_3\cr}$ transform in the same way.
The mirror symmetry $T\leftrightarrow U$ is also perturbative
in the heterotic string; it transforms
the electric charges $\hat M_I$ into electric charges and the magnetic charges
$\hat N_I$ into magnetic ones:
\beqa
\hat M_2\leftrightarrow \hat M_3,\qquad \hat N_2\leftrightarrow \hat N_3.
\label{mirror}
\eeqa
In addition, the BPS mass formula
(\ref{massphy}) is invariant under the non-perturbative $S$ duality group
$SL(2,{\bf Z})_S$ which transforms $S\rightarrow {aS-ib\over icS+d}$
and mixes the electric and magnetic charges as
\beqa
\pmatrix{\hat N_I\cr\hat M_I}\rightarrow \pmatrix{a&c\cr b&d\cr}\pmatrix
{\hat N_I\cr\hat M_I\cr}.\label{sdual}
\eeqa

As discussed in \cite{DuffLiuRahm} there is
furthermore an $S-T-U$ triality symmetry, which is related to
the string-string duality symmetries among the heterotic, type IIA and
type IIB $N=4$ four-dimensional strings.
Specifically, exchanging the $S$-field with the modulus $T$
amounts to performing the following electric magnetic duality
transformation:
\beqa
\hat M_2\leftrightarrow \hat N_0,\qquad \hat M_1\leftrightarrow \hat N_3.
\label{st}
\eeqa
This exchange corresponds to the string-string duality transformation
between the heterotic string and the type IIA string. (The
four-dimensional $N=4$ type IIA string
is obtained by compactifying the ten-dimensional IIA string
on $K_3\times T_2$.)
In the type IIA string the moduli
of $T_2$ are given by $S$ and $U$, whereas $T$
corresponds to the
string coupling constant. Thus $\hat M_1$ and $\hat M_2$ are
magnetic charges in the type IIA case, whereas $\hat N_0$ and $\hat N_3$
are
electric charges.

The transformation $S\leftrightarrow U$,
which corresponds to the string-string duality between
the heterotic and type IIB string, is obtained is an analogous way:
\beqa
\hat M_1\leftrightarrow \hat N_2,\qquad \hat M_3\leftrightarrow \hat N_0.
\label{su}
\eeqa
In the IIB string the moduli of $T_2$ correspond to $S$ and $T$, whereas
the string coupling constant is denoted by $U$.
These  two transformations
$S\leftrightarrow T$ and $S\leftrightarrow U$,
are thus of non-perturbative nature since  electric
charges and magnetic charges are exchanged.
However, as we will discuss in the following, the
exchange $S\leftrightarrow T$ is not a true symmetry of the heterotic string.
 The BPS mass spectrum of the
the heterotic (IIA, IIB) string is not symmetric under the exchange
$S\leftrightarrow T$ ($T\leftrightarrow U$, $S\leftrightarrow U$),
since the BPS masses are given by the maximum of $|Z_1|^2$ and $|Z_2|^2$.
These
operations just exchange the spectrum of the heterotic
string with the spectrum of
the type IIA, IIB strings.

\subsection{The short $N=4$ BPS multiplets}

As already said, the BPS mass formula (\ref{massphy}) is valid for
intermediate as well as for short $N=4$ supermultiplets.
Let us first consider the short $N=4$ multiplets.
Short BPS multiplets are multiplets for which
$\Delta Z^2 = |Z_1|^2 - |Z_2|^2 = 0$
at generic points in the moduli space.
Namely
$Z_1$ and $Z_2$  agree in the heterotic case
provided that the electric and magnetic charge vectors are parallel:
\beqa
\vec Q^{\rm het}||\vec P^{\rm het}.
\eeqa
That is, short multiplets
are multiplets for which $\hat{M}_I \propto \hat{N}_I$, and
the states which satisfy this constraint are characterized by
the following condition which we call the $S$-orbit or also the
heterotic orbit (a general discussion about orbits of duality groups can
be found in appendix A):
\beqa
s\hat M_I=p\hat N_I,\qquad s,p\in {\bf Z}.\label{shortm}
\eeqa
This condition can be also expressed as
\beqa
\hat M_I\hat N_J-\hat M_J\hat N_I=0.\label{equivcon}
\eeqa

Let us plug in the condition
(\ref{shortm})
into the BPS mass formula (\ref{massphy}).
Then the short multiplets have the following holomorphic masses
\cite{SchSen,Sen2}
\beqa
{\cal M} = (s + i p S) ( m_2 -i m_1U + i n_1T -n_2UT),\label{massphya}
\eeqa
where we have made the following identification:
\beqa
\hat M_0&=&sm_2,\quad \hat M_1=sn_2,\quad \hat M_2=sn_1, \quad
\hat M_3=-sm_1,
\nonumber\\
\hat N_0&=&pm_2,\quad \hat N_1=pn_2,\quad
\hat N_2=pn_1, \quad \hat N_3=-pm_1.\label{tuo}
\eeqa
We see that now the BPS masses factorize into an $S$-dependent term
and into a moduli dependent piece.
Thus for the case of
short multiplets, (\ref{massphya}) shows
that the quantum numbers $m_i$ and $n_i$
are to be thought of as the momentum and winding numbers associated
with the
2-torus parametrised by the $T,U$-moduli, whereas the quantum
numbers $s$ and $p$ denote the electric and magnetic quantum numbers
associated with the $S$-modulus. The short multiplets which fall into the
orbit (\ref{shortm}) clearly contain all elementary,
electric heterotic string states with
magnetic charge $p=0$.
For the elementary BPS states, the BPS mass is determined by the
right-moving $T^2$ lattice momentum: $M^2\sim p_R^2$; furthermore
the elementary BPS states have to satisfy $N_R+h_R=1/2$, where $N_R$ is the
right-moving oscillator number and is $h_R$ the right-moving internal conformal
dimension.
The heterotic level matching condition for elementary states reads
\beqa
{1\over 2}p_L^2-{1\over 2}p_R^2=m_1n_1+m_2n_2=N_R+h_R-N_L+{1\over 2}.
\label{levelm}
\eeqa
In the limit $S\rightarrow\infty$ an infinite number of
elementary string states with $p=0$, $s$ arbitrary become massless.
Similarly for
$S\rightarrow 0$, an infinite tower of magnetic monoples with
$s=0$, $p$ arbitray become light.

The orbit (\ref{tuo}) further decomposes into
(still reducible) suborbits
$m_1n_1+m_2n_2=a\in {\bf Z}$, as follows.

The suborbit  (i) $m_1n_1+m_2n_2=0$ contains the Kaluza-Klein excitations
of the elementary states and the Kaluza-Klein monopoles. However
this suborbit does not contain
any states which become massless for finite values of $T$ and $U$.

The second suborbit (ii)
$m_1n_1+m_2n_2=1$ contains the elementary states which become
massless within the $T,U$ moduli space.
Specifically
one gets the
following critical lines/points (modulo $T,U$ duality transformations)
(for a more detailed discussion see \cite{CLM}):

\noindent (1) $T=U$: this is the line of enhanced $SU(2)$ gauge symmetry;
the additional massless field carry the following momentum and winding numbers:
$m_1=n_1=\pm1$, $m_2=n_2=0$.

\noindent (2) $T=U=1$: here there is an enhanced $SU(2)^2$ gauge symmetry
where the four additional vector multiplets carry the charges $m_1=n_1=\pm  1$,
$m_2=n_2=0$ or $m_1=n_1=0$, $m_2=n_2=\pm 1$.

\noindent  (3) $T=U^{-1}=\rho=e^{i\pi/6}$:
this is the point of enhanced $SU(3)$
gauge symmetry with six additional massless vector multiplets
of charges $m_1=n_1=\pm 1$, $m_2=n_2=0$ or $m_1=n_1=m_2=\pm 1$, $n_2=0$
or $m_1=n_1=-n_2=\pm 1$, $m_2=0$.

In addition, this suborbit (ii) contains also the socalled $H$ monopoles.


\subsection{The intermediate $N=4$ BPS multiplets}

Let us now investigate the structure of the intermediate $N=4$ BPS
multiplets.
Intermediate BPS multiplets are multiplets for
which $\Delta Z^2 \neq 0$ at generic points
in the moduli space.  Inspection of (\ref{deltaz})
shows that intermediate multiplets are dyonic and that the vectors
$\hat{M}$ and $\hat{N}$ are not proportional to each other.
Heterotic intermediate orbits
can be characterized as follows
\beqa
\hat M_I\hat N_J-\hat M_J\hat N_I \not= 0. \label{intermorb}
\eeqa

In analogy to the constraint (\ref{tuo}) for the short heterotic multiplets
let us consider a constraint which leads to a
BPS mass formula which factorizes into a $T$-dependent and into a
$S,U$-dependent term.
Specifically
this constraint, the $T$-orbit or type IIA orbit, has the form
\beqa
\hat M_0&=&sm_2, \quad\hat M_1=-pm_1, \quad \hat M_2=pm_2, \quad\hat M_3=-sm_1,
\nonumber\\
\hat N_0&=&sn_1, \quad\hat N_1=pn_2,\quad \hat N_2=pn_1, \quad\hat N_3=sn_2,
\label{sto}
\eeqa
and the BPS mass formula (\ref{massphy}) in the heterotic case can be
written as
\beqa
{\cal M}_1 &=& (s + i p T) ( m_2 -i m_1U + i n_1S -n_2US),\nonumber\\
{\cal M}_2&=& (s+i pT)(m_2 - im_1U-in_1\bar S+n_2U\bar S).\label{massphyb}
\eeqa
This formula and the constraint
(\ref{sto}) are invariant under $SL(2,{\bf Z})_S\times SL(2,{\bf Z})_T\times
SL(2,{\bf Z})_U$.
Clearly, the constraints (\ref{sto}) and (\ref{shortm}) are just related by
the $S\leftrightarrow T$ transformation given in eq.(\ref{st}).
The states satisfying the constraint (\ref{sto}) are short\footnote{As shown
in appendix A, the short
multiplets are precisely those which are simultanously in the $T$ and in
the $S$ orbit (and therefore in the $STU$ orbit).} and
also intermediate $N=4$
multiplets in
the heterotic string theory.
However, using the string-string duality between the heterotic string and
the type IIA string, these states are short $N=4$ multiplets in the
dual type IIA theory. This means that  the orbit condition
(\ref{sto}) is satisfied for  electric and magnetic charge vectors which are
parallel
in the type IIA theory: $\vec Q^{\rm IIA}||\vec P^{\rm IIA}$
$\Leftrightarrow \hat{\bf M}^{(A)} \wedge \hat{\bf N}^{(A)} =0$.\footnote{
See the discussion given in appendix A.}
Then the transformations $SL(2,{\bf Z})_S\times
 SL(2,{\bf Z})_U\times{\bf Z}_2^{U
\leftrightarrow S}$ are perturbative in the IIA theory,
whereas $SL(2,{\bf Z})_T$
is of non-perturbative origin.
Thus, one can just repeat the analyis of
the additional massless states for the type IIA theory.
Specifically, in the type IIA theory
there is a critical line $S=U$ with two  additional massless
fields,
a critical point $S=U=1$ with four additional massless points, and
a critical point $S=U^{-1}=\rho$ with six additional massless fields.
In the case of being electric ($p=0$) these states lead
to a gauge symmetry enhancement in the
type IIA theory.
The corresponding charges immediately follow from our previous
discussion. Note, however, that the additional massless gauge bosons are
not elementary in the type II string but of solitonic nature \cite{HullTown}.

Switching again back to the heterotic theory, there are no
massless intermediate multiplets within this orbit
at the
line $S=U$ or points $S=U=1$, $S=U^{-1}=\rho$.
The reason is that we have to remind ourselves that the correct BPS masses
are given by the maximum of $|Z_1|^2$ and $|Z_2|^2$. To
illustrate this, take $S=U$
and consider
as an example the state with $p=m_2=n_2=0$, $m_1=n_1=1$ and $s$ arbitrary,
i.e. $\hat M_0=\hat M_1=\hat M_2=\hat N_1=\hat N_2=\hat N_3=0$, $\hat M_3=-s$,
$\hat N_0=s$. The BPS mass of this intermediate state
is given by $m_{\rm BPS}^2=|Z_2|^2 =
\frac{s^2}{4(T + \bar{T})}$.
  Thus we see that the heterotic BPS mass formula
is not symmetric under $S\leftrightarrow T$.

Of course, there exists another constraint, the $U$ or type IIB orbit,
\beqa
\hat M_0&=&sm_2, \quad\hat M_1=pn_1, \quad \hat M_2=sn_1, \quad\hat M_3=pm_2,
\nonumber\\
\hat N_0&=&-sm_1, \quad\hat N_1=pn_2,\quad \hat N_2=sn_2, \quad\hat N_3=-pm_1,
\label{suo}
\eeqa
for which the corresponding BPS mass formula factorises into
\beqa
{\cal M}_1& = &(s + i p U) ( m_2 -i m_1S + i n_1T -n_2ST),\nonumber\\
{\cal M}_2&=& (s+ipU)(m_2+im_1\bar S+in_1T+n_2\bar ST).
\label{massphyc}
\eeqa
The discussion of this case is completely analogous to the
previous one; the states which satisfy the constraint (\ref{suo})
correspond to the short $N=4$ BPS multiplets  in the dual type IIB theory with
$\vec Q^{\rm IIB}||\vec P^{\rm IIB}$
$\Leftrightarrow \hat{\bf M}^{(B)} \wedge \hat{\bf N}^{(B)} =0$.\footnote{
See the discussion in appendix A.}

As discussed above, the orbits (\ref{sto}) and (\ref{suo}) do not contain
additional massless intermediate states  in the heterotic theory.
There are, however, further lines in the
moduli space at which
intermediate multiplets with spin 3/2
components appear to become massless,
as it was already observed in \cite{Cvetic}.\footnote{
A massive intermediate spin 3/2 multiplet saturating one central charge
has the following component structure:
$(1\times {\rm Spin}~3/2,6\times{\rm spin}~1,
14\times{\rm spin}~1/2,14\times{\rm spin}~0$), where
the components transform as representations of
$USp(6)$. A massless spin 3/2 multiplet
has the following structure
$(1\times{\rm spin}~3/2,4\times{\rm spin}~1,(6+1)\times{\rm spin}
{}~1/2,(4+4)\times{\rm spin }~0)$.
Then, if the intermediate multiplet becomes
massless at special points in the moduli space,
the 'Higgs' effect works such that
1 massive spin 3/2 multiplet splits into a massless spin 3/2 plus
2 massless vector multiplets.} Additional massless spin 3/2 multiplets
are clearly only physically  acceptable if they lead to a consistent
enhancement
of the local $N=4$ supersymmetry to higher supergravity such as $N=5,6,8$.
However, we do not find a non-perturbative
enhancement of $N=4$ supersymmetry
at the lines of possible massless
intermediate multiplets.
Moreover
it is  absolutely not clear whether these massless spin 3/2 fields
really exist as physical soliton solutions.
In fact there are some additional good
reasons to reject these
states from the physical BPS spectrum. First the explicitly known
\cite{Cvetic} heterotic soliton
solutions for massless intermediate states are singular.
Second
an argument against the existence of massless spin 3/2 multiplets could be the
fact that such states do not exist in any fundamental string at weak coupling.
Finally, in the next chapter will argue that these kind of massless states
also do not appear in $N=2$ heterotic strings.
Nevertheless we think it is useful to further investigate the
interesting problem of non-perturbative supersymmetry enhancement in the
future. Therefore we  list the
possible massless spin 3/2 multiplets, i.e. the zeroes
of the BPS mass formula, in appendix B.

\section{The $N=2$ BPS Spectrum}

\setcounter{equation}{0}

\subsection{General formulae}

Let us now discuss the spectrum of BPS states in four-dimensional
strings with $N=2$ supersymmetry.
These masses are dermined
by the complex central charge $Z$ of the $N=2$ supersymmetry algebra:
$m_{\rm BPS}^2=|Z|^2$.
In $N=2$ supergravity the states that saturate this BPS bound
belong either to short $N=2$ hyper multiplets or to short $N=2$ vector
multiplets.
In general the mass formula as a function of
$n$ Abelian massless vector multiplets $\phi^i$
($i=1,\dots ,n_V$) is given by the following expression
\cite{FerKouLuZwi,CAFP,CAF}
\begin{equation}
m_{\rm BPS}^2=e^K|M_IP^I+ i N^IQ_I|^2=e^K|{\cal M}|^2.
\label{massnt}
\end{equation}
Here $K$ is the K\"ahler potential,
the $M_I$ ($I=0,\dots ,n_V$) are the electric quantum numbers of
the Abelian $U(1)^{n_V+1}$ gauge group and
the $N^I$ are the magnetic quantum numbers.
$\Omega =
( P^I, iQ_I)^T$ denotes a symplectic section or
period vector;
the mass formula (\ref{massnt}) is invariant under the
following symplectic $Sp(2n_V+2,{\bf Z})$ transformations, which act on the
period
vector $\Omega$ as
\begin{equation}
\pmatrix{P^I\cr i Q_I\cr}\rightarrow \Gamma\pmatrix{
P^I\cr  i Q_I\cr}=\pmatrix{U&Z\cr W&V\cr}\pmatrix{
P^I\cr i Q_I\cr},
\label{symptr}
\end{equation}
where the $(n_V+1)\times (n_V+1)$
sub-matrices $U,V,W,Z$ have to satisfy the symplectic
constraints
$
U^T V - W^T Z = V^T U - Z^T W = 1$,
$U^T W = W^T U$, $Z^T V = V^T Z$.
Thus the target space duality group $\Gamma$,
perturbatively as well non-perturbatively,  is a certain subgroup
of $Sp(2n_V+2,{\bf Z})$.

The holomorphic section $\Omega$ is determined by the vacuum expectation
values and couplings of the $n_V+1$ massless vector multiplets $X^I$ belonging
to the Abelian gauge group $U(1)^{n_V+1}$. (The
field $X^0$, which belongs to the graviphoton $U(1)$ gauge group, has
no physical scalar degree of freedom; in special coordinates it will simply
be set
to one: $X^0=1$; then one has $\phi^i=X^i$.)
Specifically, in a certain coordinate system \cite{Proy}, one can simply set
$P^I=X^I$ and the
$Q_I$  can be expressed in terms of the first derivative of an holomorphic
prepotential $F(X^I)$ which is an homogeneous function of degree two:
$Q_I=F_I={\partial F(X^I)\over\partial X^I}$. The gauge couplings as
well as the K\"ahler potential can be also expressed in terms of $F(X^I)$;
for example the K\"ahler potential
is given by
\beqa
K = - \log  ( -i \Omega^{\dag}
\left( \begin{array}{ccc}
0 & {\bf 1} \\
- {\bf 1} & 0 \\
\end{array} \right)
\Omega ) =
- \log \left( X^I \bar{F}_I + \bar{X}^I F_I \right)
\label{kp}
\eeqa
which is, like ${\cal M}$, again a symplectic invariant.

To be specific we will now consider an heterotic string which is obtained
from six dimensions as
a compactification on a two-dimensional torus $T_2$.
The corresponding physical vector fields
are defined as $S=i{X^1\over X^0}$,
$T=-i{X^2\over X^0}$, $U=-i{X^3\over X^0}$ and the graviphoton corresponds
to $X^0$. Thus there is an Abelian gauge group $U(1)^4$.

\subsection{The classical $N=2$ BPS spectrum}

Let us start by discussing the form of the classical BPS spectrum.
The classical heterotic prepotential
is given by
\cite{CAFP,WitKapLouLu,AFGNT}
\beqa
F=i{X^1X^2X^3\over X^0}=- STU.
\label{classprep}
\eeqa
This classical prepotential is obviously invariant under the full exchange
of all vector fields $S\leftrightarrow T\leftrightarrow U$.
When considering the classical gauge Lagrangian \cite{Proy}, which follows from
this prepotential, one finds a complete `democracy' among the
three fields $S$, $T$ and $U$. Specifically,
we will discuss  three types of symplectic bases (the discussion
about these bases is quite analogous to the discussion about the three $S,T,U$
orbits given in the previous chapter).

First, consider a choice of symplectic basis
(we call this the $S$-basis) in which the $S$-field plays
its conventional role as the loop counting parameter. The weak coupling limit,
i.e. the limit
when all gauge couplings become simultaneously small, is  given by
the limit $S\rightarrow\infty$.
As explained in \cite{CAFP,WitKapLouLu,AFGNT},
the period vector $(X^I,iF_I)$ ($F_I=
{\partial F\over X^I}$), that follows from the prepotential (\ref{classprep}),
does not lead to classical gauge couplings which all become small in the
limit of large $S$. Specifically, the gauge couplings which involve
the $U(1)_S$ gauge group are constant or even grow in the string weak
coupling limit $S\rightarrow\infty$ like $(S+\bar S)^{-1}$, whereas
the couplings for $U(1)_T\times U(1)_U$ behave in
the standard way as being proportional to $S+\bar S$. In order to choose a
period vector,
with all gauge
couplings being proportional to
$S+\bar S$, one has to replace $F_{\mu\nu}^S$ by its dual which is
weakly coupled in the large $S$ limit. This is achieved by the
following symplectic transformation $(X^I,iF_I)\rightarrow (P^I,iQ_I)$
where\footnote{Note however that the new coordinates $P^I$ are not
independent and hence there is no prepotential $Q(P^I)$ with the property
$Q_I={\partial Q\over\partial P^I}$.}
\be
P^1 = i F_1,\; Q_1 = i X^1,\mbox{   and   }P^i = X^i,\;Q_i = F_i\quad
{\mbox {\rm for } }\quad i =0,2,3.
\label{PQtoXF}
\eq
In the $S$-basis the classical period vector takes the form
\begin{equation}
\Omega^T=(1,TU,iT,iU,iSTU,iS,-SU,-ST),\label{clperiod}
\end{equation}
where $X^0=1$. One sees that after the transformation (\ref{PQtoXF})
all electric period fields $P^I$ depend only on $T$ and $U$, whereas
the magnetic period fields $Q_I$ are all proportional to $S$.
In this
basis $\Omega$  the holomorphic BPS masses (\ref{massnt}) become\footnote{We
call this the classical BPS spectrum, since  it is computed by using the
tree level prepotential. Nevertheless this BPS spectrum contains
non-perturbative solitons, and this formula refers to their `classical',
i.e. weak coupling, masses.}
\beqa
{\cal M}=M_0+M_1TU+iM_2T+iM_3U+iS(N_0TU+N_1+iN_2U+iN_3T)\label{holnt}
\eeqa

Let us compare these $N=2$ BPS masses with the $N=4$ BPS masses discussed
in  section 2. Specifically,
comparing with eq.(\ref{massphy}) we recognize that the
classical $N=2$ mass formula and the  $N=4$ mass formula ${\cal M}_1$
agree  upon
the trivial substitution $M_1=-\hat M_1$, $N_0=-\hat N_1$, $N_1=\hat N_0$.
(Substituting $S$ by $\bar S$ and setting $M_0=-\hat M_0$, $M_1=\hat M_1$,
$M_2=-\hat M_2$, $M_3=-\hat M_3$, $N_0=-\hat N_1$, $N_1=\hat N_0$
the $N=2$ BPS masses agree with ${\cal M}_2$.)
In contrast to $N=4$, eq.(\ref{holnt}) directly gives the correct
BPS masses without one having to take
the maximum of two in general different
central charges. The reason for this is the fact that in $N=2$ all
BPS states belong to short (vector or hyper) multiplets. In fact, when
truncating the
$N=4$ heterotic string down to $N=2$, the short as well as the intermediate
$N=4$ multiplets become short in the $N=2$ context.  This observation
potentially leads to new $N=2$ massless BPS multiplets which will
be a genuine $N=2$ effect as we discuss in the following.

The classical $U(1)^4$ gauge Lagrangian in the $S$-basis and the
classical $N=2$ BPS mass formula are invariant under the perturbative
duality symmetries  $SL(2,{\bf Z})_T\times SL(2,{\bf Z})_U
\times {\bf Z}_2^{T\leftrightarrow U}$.
As discussed above \cite{CAFP,WitKapLouLu,AFGNT},
these transformations  can be written
as specific $Sp(8,{\bf Z})$ transformations $\Gamma^{\rm classical}$
with the property
that $W^{\rm classical}=Z^{\rm classical}=0$,
${U^{\rm classical}}^TV^{\rm classical}=1$.
In addition, the field equations in the $S$-basis and
the classical BPS mass formula are also invariant
under $SL(2,{\bf Z})_S$
and, in contrast to the $N=4$ heterotic case, are also invariant
under the
transformations $S\leftrightarrow T$ and $S\leftrightarrow U$.
Of course, whether the BPS spectrum is really invariant
under the symmetries $S\leftrightarrow T$ and $S\leftrightarrow U$
depends on the non-perturbative dynamics
and cannot be read off from the
BPS mass formula. The point is that, unlike the $N=4$ case, the
$N=2$ BPS mass formula in principle allows for an $S\leftrightarrow T
\leftrightarrow U$ symmetric spectrum.
Indeed there exist some good indications
that  specific models are
$S\leftrightarrow T\leftrightarrow U$ symmetric even after
taking into account all non-perturbative
corrections. The non-perturbative
duality transformations are given by specific $Sp(8,{\bf Z})$
transformations with group elements, that have in general non-zero
submatrices $W$ and $Z$. For example,
the transformation $S\leftrightarrow T$ corresponds to the following
non-perturbative symplectic $Sp(8,{\bf Z})$ transformation:
\beqa
P^1\leftrightarrow -i Q_3,\qquad P^2\leftrightarrow iQ_1.
\label{symplst}
\eeqa
Analogously the transformation $S\leftrightarrow U$ is induced by
\beqa
P^1\leftrightarrow -iQ_2,\qquad P^3\leftrightarrow iQ_1.
\label{symplsu}
\eeqa
The symplectic transformations which correspond to $SL(2,{\bf Z})_S$ can be,
for example, found in \cite{CAFP,WitKapLouLu}.

Let us now define a second symplectic basis, the $T$-basis, in which
the $T$-field plays the role of the loop counting parameter.
In the $T$-basis all gauge couplings go to zero for large $T$. As we will
see, the $T$-basis is related to the standard $S$-basis essentially by
a non-perturbative
$S\leftrightarrow T$ transformation, i.e. by an exchange of certain
electric and magnetic fields \cite{DuffLiuRahm}.
 In exact analogy to the $S$-field
dependence of the classical gauge couplings, the prepotential
(\ref{classprep}) leads to gauge couplings of $U(1)_T$ which are constant
or grow in the limit $T\rightarrow\infty$. In order to obtain a uniform
$T$-dependence of all gauge couplings one has to perform an electric
magnetic duality transformation for $U(1)_T$, as follows
\be
\tilde P^2= i F_2,\; \tilde Q_2 = i X^2,\mbox{   and   }
\tilde P^i = X^i,\;\tilde Q_i = F_i\quad
{\mbox {\rm for } }\quad i =0,1,3.
\label{PQtoXFtil}
\eq
Then the new classical
period vector in the $T$-basis
reads $\tilde\Omega^T=(1,iS,-SU,iU,iSTU,TU,-iT,-ST)$.
We recognize that all electric periods $\tilde P^I$ do not depend on $T$,
whereas the magnetic periods $\tilde Q_I$ are propotional to $T$.
Clearly the period vector $\tilde\Omega$ is just obtained by an
$S\leftrightarrow
T$ transformation
from the period vector $\Omega$ together
with
some trivial relabeling of electric
and magnetic charges.
In the $T$-basis the classical gauge Lagrangian as well as the BPS
mass formula are invariant under the transformations $SL(2,{\bf Z})_S
\times SL(2,{\bf Z})_U\times {\bf Z}_2^{S\leftrightarrow U}$.
These transformations are of perturbative nature in the $T$-basis and
correspond to sympletic matrices $\tilde\Gamma$ with $\tilde W=\tilde
Z=0$, $\tilde U^T\tilde V=1$. On the other hand the transformations
$SL(2,{\bf Z})_T\times {\bf Z}_2^{T\leftrightarrow U}\times {\bf Z}_2^{S
\leftrightarrow T}$ are of non-perturbative nature with in general
$\tilde W,\tilde Z\neq 0$.

It is obvious that one can finally choose another period vector,
the $U$-basis,
which leads to classical gauge couplings which have a uniform dependence
on $U$ and vanish in the limit of $U\rightarrow\infty$. The corresponding
formula look analogous to the
one just discussed and can be easily written down.

Next let us discuss the form of the classical $N=2$ BPS spectrum
with special focus on the appearance of massless states.
Specifically, the singular loci of additional massless states in
the classical moduli space fall into three different classes:

\noindent
(i) First there are the elementary  states which become massless
at $T=U$, $T=U=1$ and $T=U=\rho$, for all values of $S$.
At these lines (points) the $U(1)_L^2$ gauge symmetries are classically
enhanced to $SU(2)$, $SU(2)^2$ or $SU(3)$ respectively.
In the `standard' $S$-basis the
corresponding BPS states carry only electric charges; however, when seen
in the $T,U$-basis, these states are dyonic.

\noindent
(ii) Second, suppose that the $S\leftrightarrow T\leftrightarrow U$ symmetry is
present in the BPS spectrum. Then there are
massless BPS states at the lines (points) $S=T$ \footnote{This
line was already briefly noted in reference \cite{duff2}.}, 
$S=T=1$, $S=T=\rho$
for arbitray $U$ and
analogously at $S=U$, $S=U=1$, $S=U=\rho$ for arbitray $T$.
In case of a perfect dynamical realization of the triality symmetry
$S\leftrightarrow T\leftrightarrow U$, the BPS
states are $N=2$ vectormultiplets,
and the Abelian gauge symmetries are again enhanced to $SU(2)$, $SU(2)^2$
or $SU(3)$ respectively. In the $S$-basis
these BPS states are  non-perturbative
dyons, whereas in the $T$ respectively $U$-basis these states are
purely electric.
Thus, in the $S$-basis,
 $U(1)$ factors, which are magnetic,
are enhanced at these special points in the $S,T,U$ moduli space.
The possible appearance of these
additional massless BPS fields for special values of the $S$-field is a
genuine $N=2$ effect not being possible in $N=4$.

\noindent (iii)
Third,  there are massless dyons for strong or weak coupling
$S=0$ or $S=\infty$ at the lines
eqs.(\ref{newline}) and (\ref{newliner})
and, for all $S$, at $T=U=1$, $T=U=\rho$.
These states belong to $N=2$ BPS multiplets, which
originate from $N=4$ intermediate multiplets, and are not related to
an enhancement of the Abelian gauge symmetries.
In case of a $S\leftrightarrow T$, $S\leftrightarrow U$ symmetry
there will be also analogous massless BPS states at the transformed
lines/points.

\subsection{The quantum $N=2$ BPS spectrum}

Of course, in general there will be non-perturbative corrections
which change  the classical BPS spectrum in a crucial way.
In the following we will
argue that for finite $S$, after
taking into account the non-perturbative
corrections, the classical singular lines in (i) split into lines of
massless monopoles and dyons a la Seiberg and Witten.
With respect to the massless states in (ii),
we will conjecture that in models, which are completely
$S\leftrightarrow T\leftrightarrow U$ symmetric,
there is a non-perturbative gauge symmetry enhancement for large $T$ or large
$U$.
Moreover we conjecture that for finite $T$, $U$ respectively, these lines of
massless gauge bosons are again split into lines of massless
monopoles and dyons.
 However we
will find no sign of massless states of type (iii) in the
non-perturbative spectrum.

In order to consider the form of the BPS spectrum after perturbative as well
of non-perturbative corrections,
we  make the following ansatz for the prepotential in the $S$-basis
\beqa
F =i \frac{X^1 X^2 X^3}{X^0} + (X^0)^2 \left(f^1(T,U) + f^{\rm NP}
(e^{-2\pi S},T,U)
\right)
\label{npprep}
\eeqa
Here $f^1(T,U)$ denotes the one-loop prepotential
\cite{WitKapLouLu,AFGNT} which cannot, by
simple power counting arguments, depend on $S$. Clearly, for large
$S$ one gets back the tree level prepotential. From
the prepotential (\ref{npprep}) we obtain
the following non-perturbative period vector
$\Omega^T = (P, iQ)$
\beqa
\Omega^T &=&
(1, TU - f^{\rm NP}_{S}
, iT, iU, iSTU + 2i (f^1 + f^{\rm NP})
 - iT (f^1_T + f^{\rm NP}_{T})
- iU (f^1_U +  f^{\rm NP}_{U})  \nonumber\\
&-& iS  f^{\rm NP}_{S}
, iS,
- SU + f^1_T  +   f^{\rm NP}_{T}, -ST +
f^1_U +   f^{\rm NP}_{U})\label{npperiod}
\eeqa
This leads to the following non-perturbative mass formula for the
BPS states
\beqa
{\cal M} &=& M_I P^I + i N^I Q_I=
M_0 + M_1 (TU -  f^{\rm NP}_{S}) + i M_2 T + i M_3 U
+ i N^0 (STU \nonumber\\
&+& 2 (f^1 + f^{\rm NP})
 - T (f^1_T + f^{\rm NP}_{T})
- U (f^1_U +  f^{\rm NP}_{U})  -S  f^{\rm NP}_{S})
+ i N^1 S \nonumber\\
&+& i N^2 ( i SU - i f^1_T  -i  f^{\rm NP}_{T} )
+ i N^3 ( i ST -i
f^1_U -i   f^{\rm NP}_{U})
\label{bpsmass}
\eeqa
We see that all states with $M_1\neq 0$ or $N^I\neq 0$ undergo
a non-perturbative mass shift.
We also recognize that electric states with $N^I=0$ do not get a mass shift
at the perturbative 1- loop level.
However the masses of states with magnetic charges
$N^I\neq 0$ are already shifted at the 1-loop level.

The 1-loop prepotential $f^1$ exhibits logarithmic singularities
exactly at the lines (points) of the classically enhanced gauge symmetries
and is therefore not a single valued function when transporting the moduli
fields around the singular lines
(see \cite{WitKapLouLu,AFGNT,CLM} for all the details).
For example around the singular $SU(2)$ line $T=U \neq 1,\rho$
the function $f^1$
must have the
following form \cite{WitKapLouLu,AFGNT,CLM}
\begin{equation}
f^1(T,U)={1\over \pi}(T-U)^2\log(T-U)+\Delta(T,U),
\end{equation}
where $\Delta(T,U)$ is finite and single valued at $T=U \neq 1,\rho$.
Around the point $(T,U)=(1,1)$ the
prepotential takes the form \cite{WitKapLouLu,AFGNT,CLM}
\begin{equation}
f^1(T,U=1)={1\over \pi}(T-1)\log(T-1)^2+\Delta'(T)
\end{equation}
and around $(T,U)=(\rho,\bar\rho)$ \cite{WitKapLouLu,AFGNT,CLM}
\begin{equation}
f^1(T,U={\bar \rho})={1\over \pi}(T-\rho)\log(T-\rho)^3+\Delta''(T),
\end{equation}
where $\Delta'(T)$, $\Delta''(T)$ are finite at $T=1$, $T=\rho$ respectively.
It follows that, when moving around these critical
lines via duality transformations,
one has non-trivial monodromy properties. Hence at one-loop,
the perturbative duality transformations $SL(2,{\bf Z})_T\times SL(2,{\bf Z})_U
\times {\bf Z}_2^{T\leftrightarrow U}$,
called $\Gamma^\infty$, are given in terms
of $Sp(8,{\bf Z})$ matrices with $U^\infty=U^{\rm classical}$,
$W^\infty\neq 0$, but still $Z^\infty=0$. This results in non-trivial
shifts of the $\theta$-angles at 1-loop.
In contrast to $\Gamma^{\rm classical}$, the 1-loop duality matrices
\cite{AFGNT}
do not preserve the short
orbit condition eq.(\ref{shortm}). This means that, from the $N=4$ point of
view, the $N=2$ 1-loop monodromies in general mix $N=2$ BPS states
which originate from $N=4$ vectormultiplets with hypermultiplets which are
truncated $N=4$ intermediate multiplets.

Now, taking into account the non-perturbative effects with
$e^{-2\pi S}\neq 0$ ,
the non-Abelian gauge symmetries are never restored, and
each perturbative critical line splits into two lines of massless monopoles
and dyons respectively \cite{SW1,CLM,AntoPartou}.
It follows that each semiclassical, i.e. 1-loop, monodromy  around
the lines of enhanced gauge symmetries are given by the product
of two monodromies around the singular
monopole and dyon lines, i.e. $\Gamma^\infty=\Gamma^{\rm monopole}
\times\Gamma^{\rm dyon}$
with $\Gamma^{\rm monopole},\Gamma^{\rm dyon}\in Sp(8)$ and
$W^{\rm monopole},Z^{\rm monopole},W^{\rm dyon}, Z^{\rm dyon}\neq0$.
Thus, only in the limit $S\rightarrow \infty$ is the theory still
symmetric under the perturbative duality group $SL(2,{\bf Z})_T\times
SL(2,{\bf Z})_U\times {\bf Z}_2^{T\leftrightarrow U}$.
Making an reasonable ansatz for $\Gamma^{\rm monopole}$ and
$\Gamma^{\rm dyon}$,
one can show \cite{CLM,AntoPartou}
that this splitting can be performed in such a way that
in the rigid limit one precisely recovers the results of
Seiberg and Witten. In addition the correct rigid limit was confirmed
\cite{KKLMV,AntoPartou}
by directly computing the non-perturbative monodromies $\Gamma^{\rm monopole}$
and $\Gamma^{\rm dyon}$ in type II Calabi-Yau compactifications
with $h_{11}=2$, i.e. for models with two vector fields $S$ and $T$.

Consider for example the splitting of the critical
line $T=U$ with
classically enhanced gauge group $SU(2)$;
the associated magnetic monopole  has non vanishing magnetic
quantum numbers $N^3=-N^2$.  Like the
massless gauge bosons before, this magnetic monopole corresponds to
a short $N=4$ vector multiplet, i.e. it
belongs to the first orbit (\ref{tuo}). Using (\ref{bpsmass})
its mass vanishes for $Q_2=Q_3$, which leads to following singular monopole
locus
\beqa
i S(T-U) - i (f^1_T - f^1_U) - i ( f^{\rm NP}_T -  f^{\rm NP}_U ) =0
\label{locusmw1}
\eeqa
Similarly, the locus of
massless dyons with charges $M_2=-M_3=N^3$, $N^2=-N^3$
has the form $T-U=Q_2-Q_3$.
Like
$\Gamma^\infty$, $\Gamma^{\rm monopole}$ and
$\Gamma^{\rm dyon}$ do  not preserve the heterotic short
orbit condition eq.(\ref{shortm}).

Let us now suppose that the full non-perturbative theory is symmetric
under the exchange symmetry $S\leftrightarrow T$. In fact the existence
of this type of quantum symmetry was already observed in models with only
two fields $S$ and $T$ \cite{KLM,KKLMV,AntoPartou}.
If this symmetry is exact we expect that in the
`weak
coupling limit' $T\rightarrow\infty$ one finds an enhancement of the Abelian
gauge group at special points in the $S,U$ moduli space.
Specifically, at $S=U$ the enhanced gauge group should be $SU(2)$,
at $S=U=1$ one has $SU(2)^2$ and at $S=U^{-1}=\rho$ one should find $SU(3)$.
In the limit
$T\rightarrow\infty$ the non-perturbative prepotential, written in
the symplectic $T$-basis, then takes the form
\begin{equation}
f(S,U)={1\over \pi}(S-U)^2\log(S-U)+\dots ;
\end{equation}
at the point $(S,U)=(1,1)$ the
prepotential takes the form
\begin{equation}
f(S,U=1)={1\over \pi}(S-1)\log(S-1)^2+\dots
\end{equation}
and around $(S,U)=(\rho,\bar\rho)$
\begin{equation}
f(S,U={\bar \rho})={1\over \pi}(S-\rho)\log(S-\rho)^3+\dots .
\end{equation}
It follows that, when moving around these critical
lines via duality transformations,
one has non-trivial monodromy properties just like
at one loop for large $S$.  At large $T$ the theory is symmetric under the
duality transformations $SL(2,{\bf Z})_S
\times SL(2,{\bf Z})_U\times {\bf Z}_2^{S
\leftrightarrow U}$ , called $\tilde\Gamma^\infty$,
which are then given in terms
of $Sp(8,{\bf Z})$ matrices with $\tilde U^\infty\tilde V^T=1$,
$\tilde W^\infty\neq 0$, $\tilde Z^\infty=0$.

What will happen if we turn on the coupling $e^{- 2 \pi T}$
in a $S\leftrightarrow
T$ symmetric theory? In the spirit of Seiberg and Witten we expect that
the lines of enhanced gauge symmetries at $T=\infty$ again split
into two lines of massless monopoles and dyons for
finite $e^{- 2 \pi T}$. The corresponding monopole
and dyon monodromies $\tilde \Gamma^{\rm monopole}$
and $\tilde \Gamma^{\rm dyon}$ are just given by conjugating
$\Gamma^{\rm monopole}$ and
$\Gamma^{\rm dyon}$ by the generator of the $S\leftrightarrow T$ exchange
symmetry.
An analogous discussion of course applies for the non-perturbative
symmetry $S\leftrightarrow U$.

In order to make the existence of the $S\leftrightarrow T$,
$S\leftrightarrow U$  symmetries
in certain type of models more plausibel
it is very useful to
utilize the (conjectured, however already quite
well established) duality
\cite{KacVaf,KLM,AntoPartou,KKLMV,FerHarvStrVa,KaLou,AldFoIbaQue,AspLou}
between heterotic $N=2$ strings and
type II $N=2$ strings on a suitably choosen Calabi-Yau backgrounds.
Specifically  consider a Calabi-Yau background characterized by the
two Hodge numbers $h_{11}$ and $h_{21}$. In the type IIA models,
$h_{11}$ must agree with the number of massless vector multiplets $n_V$
in the heterotic model.
The number of hypermultiplets $n_H$ is given by $h_{21}+1$
where the extra 1 accounts for the type II dilaton.
Since the type II dilaton sits in an $N=2$ hypermultiplet
and does not couple to the vector multiplets,
the  classical type II prepotential is exact.
It follows that  BPS spectrum of the form (\ref{massnt})
is exact in the type II case as well.  For the type IIA
models, the Calabi-Yau
world-sheet instanton effects  then correspond to the
target space instanton effects on the heterotic side \cite{FerHarvStrVa}.

After performing the mirror map
from IIA to IIB, the number of massless fields in
the IIB Calabi-Yau compactification is determined as $n_H=h_{11}+1$,
$n_V=h_{21}$.
In the type IIB case the holomorphic prepotential receives no world
sheet instanton corrections; it
becomes singular at
the socalled conifold points in the Calabi-Yau moduli space and at some
other isolated points.
Then, within the string-string duality picture the type IIB singular
locus
just corresponds to the
locus of massless magnetic monopoles or dyons on the heterotic side (see the
discussion below).

Let us first consider as the most simple example the case  with only two vector
fields $S$ and $T$. Specifically we  consider the Calabi-Yau
space, constructed as a
hypersurface of degree 12 in $WP_{1,1,2,2,6}(12)$,
with $h_{11}=2$ and $h_{21}=128$ \cite{CanOsFoKaMo}. It
was observed in \cite{KLM,KKLMV} that this model
indeed possesses an exchange symmetry $S\leftrightarrow T$ at the
non-perturbative level.  This symmetry can be recognized by looking at the
instanton expansions, as done in \cite{KLM}.
Specifically, the transformation $q_1\rightarrow q_1q_2,q_2\rightarrow 1/q_2$
($q_1=e^{i 2 \pi t_1}=e^{- 2 \pi T}$,
$q_2=e^{i 2 \pi t_2}=e^{- 2 \pi (S-T)}$)
can be traced back to the monodromy considerations of \cite{CanOsFoKaMo}.
Under this transformation
$
n_{j,k}q_1^jq_2^k\rightarrow n_{j,k}q_1^jq_2^{j-k}
$,
so that the non-perturbative symmetry should come from
$
n_{j,k}=n_{j,j-k}
$
where the $n_{j,k}$ are world-sheet instanton numbers of genus zero.
Indeed, it was shown in
\cite{CanOsFoKaMo} (there in the model
$P_{1,1,2,2,2}(8)$, but this makes no difference here)
that the homology type of the holomorphic image of the worldsheet
$\Sigma$ changes as
$
\Sigma_{j,k}=jh+kl\rightarrow j(h+l)+k(-l)=j' h+k' l
$
with
$
j'=j$, $k'=j-k
$
under the monodromy
$
T_{\infty}:(t_1,t_2)\rightarrow (t_1+t_2,-t_2+1)
$
on the periods $(t_1,t_2)$.

The singular discriminant locus
of this Calabi-Yau, on which certain BPS
states become massless, is given by the
following equation \cite{KacVaf}
\beqa
\Delta=(1-y)((1-x)^2-x^2y).\label{locussimple}
\eeqa
The conventional weak coupling limit is given by
$y=e^{- 2 \pi S_{\rm inv}}= 0$.
In this limit  one recovers the perturbative duality symmetry
$SL(2,{\bf Z})_T$
($S_{\rm inv}$ is the 1-loop redefined $S$-field, invariant
under the perturbative duality group), and
the parameter $x$ can be expressed \cite{KLM,KKLMV}
in terms of modular functions as $x=1728/j(T)$.
In the limit $y=0$, $\Delta$ degenerates into the quadratic factor $(1-x)^2$,
and at $x=1$, i.e. $T=1$, one finds the classical $SU(2)$ gauge symmetry
enhancement. For $y\neq 0$ this line splits into the two lines of massless
monopoles and dyons \cite{KLM,KKLMV}.

Using the $S\leftrightarrow T$ symmetry,
the discriminant locus  has
a second `weak coupling' limit, where $\Delta$ quadratically degenerates.
We suggest to identify this limit with $T\rightarrow\infty$;
in this limit $SL(2,{\bf Z})_S$ should be a symmetry of the theory.
In the limit $T\rightarrow\infty$ the coupling constant $y$ should
be given as $y=e^{-2\pi T_{\rm inv}}\rightarrow 0$, where $T_{\rm inv}$ is
a redefined modulus, invariant under $SL(2,{\bf Z})_S$. 
Then $\Delta$ takes again the form $\Delta\sim(1-x)^2$ which signals a $SU(2)$
gauge symmetry enhancement at $x=1$ now corresponding to $S=1$. 
Thus we conjecture to make the
following identification for large $T$: $x=1728/j(S)$. Observe that for
large $S$, this $x$ is exponential in $S$: $x\rightarrow e^{- 2 \pi S}$.
Turning on the coupling $y$, the quadratic degeneracy
is again lifted, and we expect that the large $T$ gauge group enhancement
is replaced by the existence of a massless monopole, dyon pair.
Recall that in the weak coupling limit $S\rightarrow\infty$ 
the appearance of the
modular function $j(T)$ originates from the fact that the underlying
Calabi-Yau space can be constructed as a $K_3$-fibration \cite{KLM}, where
$S$ plays the role of the size of the base space \cite{AspLou}.
In analogy,  the $S\leftrightarrow T$ symmetric picture
could then mean that there exist a dual `quantum $K_3$ fibration'
with $T$ being the modulus of the base space, implying
the appearance of the
modular function $j(S)$ in the limit $T\rightarrow\infty$.

Now let us investigate the case of three moduli $S,T,U$.
The singular loci of massless BPS states for this type of $N=2$ string
models
was recently derived \cite{KKLMV}
from a type IIB compactification
on a Calabi-Yau space $WP_{1,1,2,8,12}(24)$ with $h_{21}=3$
and $h_{11}=243$ leading to 244 hypermultiplets (including
the type II dilaton multiplet).
In ref.\cite{KLM} some arguments
were given supporting the conjecture that this model is symmetric
under the exchange $S\leftrightarrow T$, $S\leftrightarrow U$.
The discriminant locus of the  $P_{1,1,2,8,12}(24)$ Calabi-Yau is
\cite{KacVaf,KLM}
\begin{equation}
\Delta=(y-1)\times {(1-z)^2-yz^2\over z^2}\times
{((1-x)^2-z)^2-yz^2\over z^2}=
\Delta_y\times \Delta_z\times \Delta_x.\label{disclocus}
\end{equation}
$x,y,z$ are functions of the three vector fields $S,T,U$.

Let us now  consider three differents limits where $\Delta$ degenerates
into quadratic factors that signal an enhancement of the Abelian gauge
symmetry at special (boundary) points in the moduli space.

\noindent (i) First consider the conventional
classical limit $y=e^{- 2 \pi S_{\rm inv}}=0$.
In this limit  one recovers the perturbative duality symmetry
$SL(2,{\bf Z})_T\times SL(2,{\bf Z})_U\times {\bf Z}_2^{T\leftrightarrow U}$
(again, $S_{\rm inv}$ is the 1-loop redefined $S$-field, invariant
under the perturbative duality group), and
the parameters $x,z$ can be expressed
in terms of the fields $T,U$ as follows \cite{KLM,LianYau,KKLMV}:
\beqa
x &=& {1\over 864}{j(T)j(U)+\sqrt{j(T)j(U)(j(T)-1728)(j(U)-1728)}\over
j(T)+j(U)-1728},\nonumber\\
z&=&864^2{x^2\over j(T)j(U)}.\label{xzfunc}
\eeqa
In this limit the two equations
${\Delta_x}=0$ and ${\Delta_z}=0$ are completely equivalent.
They both correspond to the classical enhancement
of one Abelian $U(1)$ gauge group to $SU(2)$. Both equations are solved
only by the
relation $j(T)=j(U)$, the line of enhanced $SU(2)$ gauge symmetry.
More exactly, $\Delta_x$ and $\Delta_z$ are double valued functions in terms
of $j(T),j(U)$. For $j(T)=j(U)$ the branch points are at $j(T)=0$ and
$j(T)=1728$, i.e. at $T=\rho$, $T=1$ respectively and
at all the points obtained by duality
transformations of these two points. With $j(T)=j(U)$
one obtains in the first branch that $z=1$, $x={1\over 864}j(T)$,
${\sqrt\Delta_x}={4j(T)(j(T)-1728)\over 1728^2}$, ${\sqrt\Delta_z
}={(j(T)-j(U))^2\over 4j(T)(j(T)-1728)}=0$. The points $x=0$, $x=2$,
where $\Delta$ further degenerates, correspond to the points
of enhanced gauge symmetries $SU(3)$ or $SU(2)^2$ respectively.
In the second branch $j(T)=j(U)$ belongs to
$z={1728^2\over(2j(T)-1728)^2}
$, $x=1\pm\sqrt z$,
${\sqrt\Delta_x}={(j(T)-j(U))^2\over 4j(T)(j(T)-1728)}=0$,
${\sqrt\Delta_z}={4j(T)(j(T)-1728)\over 1728^2}$.
However the product ${\sqrt{\Delta_x\Delta_z}}$ is single valued,
and one obtains as an identity
in the limit $y=0$ \cite{Cu}: ${\sqrt\Delta_x\sqrt\Delta_z}
={(j(T)-j(U))^2
\over 1728^2}$. In summary,
in the classical limit $y=0$ one precisely finds the lines (points)
of enhanced gauge symmetries,
 namely first $SU(2)$ with $j(T)=j(U)$
corresponding to $T=U$ (plus all
dual equivalent lines), second $SU(2)^2$ with
$j(T)=j(U)=1728$ corresponding to $T=U=1$ and third $SU(3)$
with $j(T)=j(U)=0$ corresponding to $T=U=\rho$.

\noindent (ii) There exists a second limit where $\Delta$ degenerates into
quadratic factors. We conjecture that this limit corresponds
to $T\rightarrow\infty$ and make  in this limit the identification
$y=e^{- 2 \pi T_{\rm inv}}$. In this limit the theory is invariant under
$SL(2,{\bf Z})_S\times SL(2,{\bf Z})_U\times {\bf Z}_2^{S\leftrightarrow U}$
and $T_{\rm inv}$ is a redefined modulus, invariant under this group.
Thus for large $T$,
$\Delta=0$ at the line $z=1$ which should be the line of
enhanced $SU(2)$ gauge symmetry for $S=U$.
Analogous to the previous case one should get a further
degeneration at the two points $S=U=1$ and $S=U=\rho$,
with enhanced  gauge groups  $SU(2)^2$, $SU(3)$
respectively.
For $T\neq\infty$, the quadratic degeneracy is lifted, and we expect
that the solutions of $\Delta=0$ correspond to lines of massless monoples and
dyons.
Unfortunately we are at the moment not ready to prove all
these conjectures. It would require a complete reorganisation of
the instanton sums in the type II mirror map.

Finally there should 
exist also a third quadratic degeneration of $\Delta$, namely
 in the
limit $U\rightarrow\infty$ 
 with $SL(2,{\bf Z})_S\times SL(2,{\bf Z})_T\times {\bf Z}_2^{S
\leftrightarrow T}$ duality symmetry. In this limit the
gauge symmetry enhancement then takes place at $S=T$, $S=T=1$ and
$S=T=\rho$.

At the end of this section let us also mention that in the quantum case we did
not find any trace of
those massless
states which we discussed under point (iii) at the classical level.
If they would exist they should have shown up for large $S$ in
$\Delta$, since they
were classically present for any $S$ and hence in particular for weak
coupling. 
 This observation
may be one more argument against the existence of the corresponding
massless intermediate states in the $N=4$ heterotic string.

\section{$N=4$ BPS sums}

\setcounter{equation}{0}

\subsection{The $N=4$ free energy}

In the next sections we will discuss the
topological string partition function
as a sum over BPS states. A similar type of partition function was
introduced in \cite{FerKouLuZwi}.
More recently the sum over BPS states was also discussed by Vafa in
\cite{Vafa}.
Concretely, let us define the following partition function $Z$\footnote{Here,
$Z$ is not to be confused with the central charge.}
\beqa
\log Z=\sum_{\rm BPS~~states}\log m_{\rm BPS}^2\label{freeen}
\eeqa

In the following, we will discuss the non-perturbative partition function
obtained
by summing over the heterotic $N=4$ BPS spectrum.
Specifically, we will consider
the following holomorphic free energy
\beqa
{\cal F}=\sum_{\hat M_I,\hat N_I}\log{\cal M}_{1,2},\label{nfourf}
\eeqa
where the holomorphic BPS masses are given in (\ref{massphy}).
The holomorphic free energy and the non-holomorphic partition function are
related as
\beqa
Z=e^{{\cal F}+\bar{\cal F}}e^K,\label{partit}
\eeqa
where $K$ is the K\"ahler potential of the moduli fields $S,T,U$ (we
are restricting the discussion to an $SO(2,2)$-coset subspace of the
toroidal moduli space).
$K$ is given by
\beqa
K=-\log\lbrack (S+\bar S)(T+\bar T)(U+\bar U)\rbrack,
\eeqa
which transforms under $SL(2,{\bf Z})_S$, $S\rightarrow {aS-ib\over
icS+d}$, as $K\rightarrow K+\log(icS+d)+\log(-ic\bar S+d)$
and likewise for $SL(2,{\bf Z})_T$ and $SL(2,{\bf Z})_U$.
It follows that $e^{{\cal F}}$ must be a modular function of modular weight
-1 under $SL(2,{\bf Z})_S$, $SL(2,{\bf Z})_T$ and $SL(2,{\bf Z})_U$
in order for $Z$ being completely duality invariant.
${\cal F}$ and $Z$  are clearly non-perturbative expressions
since they involve the summation over elementary string states as well as
over soliton states like magnetic monopoles etc.
Thus $Z$, ${\cal F}$ will exhibit
the non-perturbative dilaton dependence of the string partition function.
Note that by demanding $Z$ to be completely duality invariant we are
requiring
the absence of non-perturbative duality anomalies, in particular the
absence of $S$-duality anomalies.

The sum eq.(\ref{nfourf}) can be more conveniently computed by selecting
some specific  summation
orbits. One criterion of selecting the
relevant summation orbits is that at least all singularities
of the free energy have to be contained in the correct way; in other words,
this means that the sum has to contain all possible states which can become
massless at certain points in the moduli space. In addition, the duality
invariance
of the free energy must not be destroyed by summing over specific orbits.
Let us start  by first summing over the three
orbits (\ref{tuo}), (\ref{sto}) and (\ref{suo}),
which are related by the triality
exchange $S\leftrightarrow T\leftrightarrow U$. (Each of these summation
orbits will contain further suborbits.)
${\cal F}_{T\leftrightarrow U}$ sums over the BPS states in the first orbit
(\ref{tuo}) and is invariant under $T\leftrightarrow U$.
Therefore ${\cal F}_{T\leftrightarrow U}$ is summing over the short
heterotic $N=4$
vector multiplets; using  eq.(\ref{massphya})
${\cal F}_{T\leftrightarrow U}$ becomes
\be
{\cal F}_{T \leftrightarrow U} = \sum_{ \{\hat M_I, \hat N_I | C_{KL} = 0 \} }
\log ( \hat M_0 - \hat M_1 TU + i \hat M_2 T + i \hat M_3 U
+ i \hat N_0 S -i \hat N_1 STU
- \hat N_2 ST - \hat N_3 SU )
\eq
In order to perform the sum one must solve the six equations
$C_{KL} =0$ in terms of unconstrained summation variables.
This can be done generalizing a method described in
\cite{OoguriVafa}. Consider the three equations $C_{0i} = 0$ first.
Setting $\hat N_1 = \hat N_2 = \hat N_3 =0$ they are fulfilled if either
(i) $\hat N_0 = 0$ or (ii) $\hat M_1 = \hat M_2 = \hat M_3 =0$.
In case (ii) the other three equations are already solved. It only
remains to sum over two unconstrained variables $s := \hat M_0$ and
$p := \hat N_0$. Thus the first contribution to the sum is
$\sum_{(s,p) \not= (0,0)} \log (s + i p S)$.
In case (i) we are left with four unconstrained variables
$m_2 := \hat M_0$, $n_2 := \hat M_1$, $n_1 := \hat M_2$ and $m_1 := - \hat M_3$
resulting in a second contribution
$\sum_{(m_1, m_2, n_1, n_2) \not= (0,0,0,0)}$
$\log (m_2 - i m_1 U + i n_1 T - n_2 TU)$. Summarizing we have
succeded in writing ${\cal F}_{T \leftrightarrow U}$ as an
unconstrained sum
\be
{\cal F}_{T \leftrightarrow U} =
\sum_{(s,p) \not= (0,0)} \log (s + i p S)  +
\sum_{(m_1, m_2, n_1, n_2) \not= (0,0,0,0)}
\log (m_2 - i m_1 U + i n_1 T - n_2 TU)
\label{ftu12}
\eq
which splits into a non--perturbative part, which only involves $S$,
and into a perturbative part only depending on the moduli $T$ and $U$.

Consider the first term in (\ref{ftu12}).  The regularized
sum \cite{OoguriVafa,FerKouLuZwi}
over the electric and magnetic charges $s,p$ leads
to the following contribution: $\sum_{s,p}\log(s+ipS)=
\log\eta(S)^{-2}$, where $\eta$ is the Dedekind function. This term
describes the non-perturbative $S$ dependence of ${\cal F}_{
T\leftrightarrow U}$. $e^{\cal F}_{T\leftrightarrow U}$
transforms as a modular function of modular weight -1 under
$SL(2,{\bf Z})_S$. ${\cal F}_{T\leftrightarrow U}$
diverges linearly for large $S$ as
well as for small $S$. These divergences reflect the appearance
of infinitely many massless electric or magnetic states for
$S\rightarrow\infty,0$ respectively.
Now, in order to evaluate the second term in expression (\ref{ftu12})
we split the
sum into the two further
suborbits, namely (i): $m_1n_1+m_2n_2=0$, (ii): $m_1n_1+m_2n_2=1$.
This choice is dictated by the appearance of the massless fields (see
\cite{CaLuMo}
for a detailled discussion). The suborbit (i) contains no states which become
massless for finite $T$ and $U$ but infinitely many states (Kaluza-Klein and
winding modes) which become massless in the degeration limits $T,U\rightarrow
0,\infty$.
Summing over the suborbit (i) leads to a term $\log\eta(T)^{-2}\eta(U)^{-2}$
with linear divergences for $T\rightarrow \infty,0$ due
to the massless Kaluza Klein states or winding modes in this limit.
The second suborbit (ii) contains the finite number of states
which are massless at the critical points in the moduli space $T=U$, $T=U=1$
and $T=U=\rho$. Then
the suborbit (ii) leads to \cite{CaLuMo} $\log(j(T)-j(U)$,\footnote{Here
we have assumed that the regularization procedure is modular
invariant. This assumption however may not hold, and the regularization
procedure
may possess
a kind of modular anomaly which
destroys the duality covariance of the sum; thus non-modular invariant, but
completely finite terms may be added to the regularized sum. These finite terms
can be absorbed by a redefinition of the dilaton field
\cite{WitKapLouLu}.} where $j$ is the
absolute modular invariant function.
This expression is
logarithmically divergent at the critical lines (points) where the
residues of the poles correctly agree with the number of massless
fields at the symmetry enhancement points.
Thus collecting the different terms (the higher orbits $m_1n_1+n_2m_2>1$ do
not give new terms)
we obtain the following holomorphic free energy
\beqa
{\cal F}_{T\leftrightarrow U}=\log(\eta(S)^{-2}\eta(T)^{-2}\eta(U)^{-2}
(j(T)-j(U))^r).\label{etas}
\eeqa
The coefficient $r$ is undetermined at this stage and corresponds
to the overall number of states becoming massless at
the specific lines (points).
Clearly $e^{{\cal F}_{T\leftrightarrow U}}$ transform as a modular function of
modular weight -1 under $SL(2,{\bf Z})_S \times SL(2,{\bf Z})_T\times
SL(2,{\bf Z})_U$, and it is invariant under $T\leftrightarrow U$ (up to
a possible extra overall $\pm$ sign).

Next let us discuss the  sum ${\cal F}_{S\leftrightarrow U}$ over the
orbit eq.(\ref{sto}).
At the first glimpse one could believe that this sum is just obtained by
performing $S\leftrightarrow T$ exchange in ${\cal F}_{T\leftrightarrow U}$.
This conclusion would be true, if there were
intermediate massless states in the second
orbit for $S=U$, $S=U=1$, $S=U=\rho$ in the heterotic
string theory. They however do not exist. Thus we conclude that the
suborbit (ii) does not lead to singularities  for finite $S,T,U$.
The fact that ${\cal F}_{T\leftrightarrow U}$ and
${\cal F}_{S\leftrightarrow U}$
do not agree reflects the non-invariance of the heterotic BPS spectrum
under the exchange $S\leftrightarrow T$.

Finally, the discussion about
the sum over the orbit eq.(\ref{suo})
is completely
analogous to the previous case.

In case that there exist massless intermediate states
at specific points/lines
in the moduli space, these states would also contribute to
the free energy. However, as we have discussed in section 2.3
there are
many good reasons to discard these massless spin 3/2 BPS soliton states.
Thus we take eq.(\ref{etas}) as the complete result for the
$N=4$ heterotic free energy.
The associated  non-perturbative partition function is invariant
under $SL(2,{\bf Z})_S\times SL(2,{\bf Z})_T\times  SL(2,{\bf Z})_U \times
{\bf Z}_2^{T\leftrightarrow U}$.
It is very similar to the ordinary bosonic string
partition function. The type IIA (IIB) partition function is
finally obtained by the exchange $S\leftrightarrow T$ ($
S\leftrightarrow U$) in eq.(\ref{etas}).

\subsection{Absence of $N=4$ thresholds and
the role of the $N=4$ free energy}

In the $N=4$ case the free energy does not correspond to
threshold corrections in the low energy effective action, since
loop corrections are absent in $N=4$, even at the non-perturbative level.
In the following we will, for example, first recall the absence of
1-loop gravitational
threshold corrections in $N=4$ heterotic strings.

In $N=4$ compactifications of the heterotic string the dilaton
$S=\frac{1}{g^2} - i \frac{\theta}{8\pi^2}$
parametrises a K\"{a}hlerian $SU(1,1)$-coset, whereas the
non-K\"{a}hlerian $SO(6,22)$-coset is parametrised by moduli
$\Phi^I$.

In a string calculation, 1-loop corrections to gravitational couplings
in $N=4$ heterotic compactifications should, if present, be of the
form
\beqa
\frac{1}{g^2_{grav}} = 12 (S+\bar{S}) + \frac{b_{grav}}{16 \pi^2} \log
\frac{M^2_{string}}{p^2} + \Delta(\Phi^I)
\label{g1string}
\eeqa
$\Delta(\Phi)$ denotes the moduli dependent
1-loop corrections due to both massless
and massive modes in the theory.  $b_{grav}$, on the other hand,
denotes the gravitational beta function coefficient computed
from the massless fields.  Note that the scale appearing
in the logarithm in (\ref{g1string}) is the string scale, as it
should for a string calculation.

In a field theory calculation, on the other hand, it is the
Planck scale which should appear in a 1-loop calculation. Thus,
consider rewriting (\ref{g1string}) as
\beqa
\frac{1}{g^2_1} = 12 (S + \bar{S}) + \frac{b_{grav}}{16 \pi^2} (\log
\frac{M^2_{Planck}}{p^2} + K) + \Delta(\Phi^I)
\label{g1}
\eeqa
where $K=- \log(S + \bar{S})$.  Here $K$ denotes the K\"{a}hler potential
for the K\"{a}hlerian $SU(1,1)$-coset parametrised by the dilaton field $S$.
We have used that $M^2_{Planck} \propto (S+\bar{S}) M^2_{string}$.
Actually, a field theory calculation would a priori give that
\beqa
\frac{1}{g^2_1} = 12(S + \bar{S}) + \frac{b_{grav}}{16 \pi^2} \log
\frac{M^2_{Planck}}{p^2} + \frac{c_{grav}}{16 \pi^2} K + \Delta(\Phi^I)
\label{g1field}
\eeqa
with some coefficient $c_{grav}$.
The term proportional to
$\frac{b_{grav}}{16 \pi^2} \log
\frac{M^2_{Planck}}{p^2}$ arises from a 1-loop graph with 2 external
gravitational legs sticking out and massless fields running in the loop.
The term proportional to $c_{grav}K$ arises from a triangle graph with
2 gravitational legs and one $a_m$ leg sticking out and with massless
fields running in the loop.  Indeed, as shown in \cite{Roo}, every
fermion in the $N=4$ theory couples to the "K\"{a}hler connection"
$a_m \propto \partial_S K \; \partial_m S - c.c$ associated to the
$SU(1,1)$-coset (note again that the $SO(6,22)$-coset is not K\"{a}hlerian
and hence there is no "K\"{a}hler connection" associated to it).
If the field theory calculation
is to match the string calculation (\ref{g1string}), then
one has to find that $b_{grav}=c_{grav}$ in the field theory calculation.

Consider now calculating $b_{grav}$ and $c_{grav}$ in field theory.
$b_{grav}$ is nothing but the sum over the trace anomalies of the
massless multiplets in the theory.  At generic points in the
$SO(6,22)$-moduli space,
the massless multiplets around are the $N=4$ supergravity multiplet
and 22 abelian $N=4$ vector multiplets.
The trace anomaly for an $N=4$ vector multiplet is zero, as
it is wellknown.  What about
the trace anomaly of the $N=4$ supergravity multiplet?
For an $N=4$ compactification of the heterotic string, the axion is
not really a scalar degree of freedom but rather an antisymmetric tensor
degree of freedom.\footnote{The associated $N=4$ supergravity multiplet
will thus contain 1 graviton, 4 gravitini, 6 graviphotons, 4 Weyl fermions,
1 antisymmetric tensor and one real scalar.}
Taking into account the following trace anomaly
contributions (in units where a real scalar degree of freedom contributes
an amount of 1) \cite{CrisDuff}, namely
$1$ from a real scalar field, $\frac{7}{4}$
from a Weyl fermion, $-13$ from a vector field, $212$ from a graviton,
$- \frac{233}{4}$ from a gravitino and $91$ from an
antisymmetric  tensor,
then gives that $b_{grav}=0$ for an $N=4$ heterotic compactification.

Since it must be that $b_{grav}=c_{grav}$, it follows that one should
for consistency also find that $c_{grav}=0$ in a field theory
calculation.  $c_{grav}$ is nothing but the K\"{a}hler
anomaly coefficent.  Using the $N=1$ assignments for the K\"{a}hler
charges one has that the fermions in the $N=4$ gravitational
multiplet carry charges $+1$, whereas the gauginos in the $N=4$
vector multiplets carry charges $-1$.
Then it follows that indeed $c_{grav}=4(21 + 1 - 22)=0$.

The fact that $b_{grav}=c_{grav}=0$
indicates that there are no
1-loop corrections to $g^2_{grav}$ in $N=4$ heterotic compactifications
at all, as indeed shown by string scattering amplitude
calculations in the context of
orbifold compactifications \cite{AnGaNa}.

Thus,
the $N=4$ holomorphic free energy cannot correspond to threshold
corrections in the low energy effective action.
What role then does the $N=4$ non-holomorphic free energy discussed in the
previous section play in the context of $N=4$ heterotic strings?
We conjecture that it is the S-duality invariant
partition function of topologically twisted $N=4$ heterotic
string compactifications.  A priori one might expect the
partition function of the topologically twisted
theory to be holomorphic in the moduli fields.
However, it was pointed out in \cite{Witten/Vafa} that, at least
in the context
of topologically twisted
$N=4$ super Yang-Mills theory on four-manifolds,
there are examples
where this is not the case due to the appearence of holomorphic
anomalies.  Hence, it is possible that
the non-holomorphicity
of the $N=4$ free energy is again a manifestation of the appearance
of holomorphic anomalies in the twisted version of $N=4$ string
compactifications.

If indeed the $N=4$ free energy is to be identified with the
partition function of
topologically twisted $N=4$
string compactifications, then this implies that,
whereas the holomorphic
gravitational coupling ${\cal F}_{grav} = 24 S$ of the untwisted
model doesn't receive perturbative or non-perturbative corrections,
the holomorphic coupling ${\cal F}_1$ of the twisted model
is more complicated and given by (\ref{etas}).
Something similar happens in the
case of twisted $N=4$ super Yang-Mills theory on four-manifolds.
There, it was found \cite{Witten/Vafa} that
S-duality invariance of the twisted partition function
only holds provided that
there
are certain non-minimal couplings in the Lagrangian
of the form $\log \eta(S) \chi$
that involve the
background gravitational field, where
$\chi$ denotes the Euler characteristic of the four-manifold
($\chi \propto \int GB$, where $GB$
denotes the Gauss-Bonnet combination).
Namely, the partition function $Z[S]$ for the topologically twisted
$N=4$ super Yang-Mills theory transforms like a modular form
with modular weight $w$ under
$S \rightarrow \frac{1}{S}$
\beqa
Z[S] \rightarrow S^{w \chi}
Z[S]
\eeqa
(ignoring the issue of holomorphic anomalies).
The following modified partition function
\beqa
{\hat Z}[S] = e^{-{\cal F}_1 \chi} Z[S]
\eeqa
however, is invariant under $S \rightarrow \frac{1}{S}$
provided that ${\cal F}_1 \propto \log \eta(S)$.

\section{$N=2$ BPS sums}

\setcounter{equation}{0}

\subsection{The $N=2$ free energy}

Let us again define the $N=2$ holomorphic free energy ${\cal F}$ as the sum
over the $N=2$ BPS states (\ref{massnt}), that is
\beqa
{\cal F}=\sum_{M_I,N^I} \log (M_IP^I+iN^IQ_I).\label{ntwofree}
\eeqa
This formula was  introduced in \cite{FerKouLuZwi} in the context of
string compactifications on Calabi-Yau spaces.
Like in the previous $N=4$ case it is useful to split this
sum into sums over the different orbits of the relevant duality
group $\Gamma$.
Since the $N=2$ K\"ahler potential changes under duality
transformations $\Gamma=\pmatrix{U&Z\cr W&V}$ as
\beqa
K\rightarrow K+\log|U^0_IP^I/P^0|^2\label{kptransf}
\eeqa
the holomorphic $N=2$ free has to transform as
\beqa
{\cal F}\rightarrow {\cal F}-\log U^0_IP_I/P^0. \label{ntwoftrans}
\eeqa

The non-perturbative heterotic $N=2$ free energy based on the
non-perturbative BPS mass formula (\ref{bpsmass})
is in general very difficult to compute. It is clear that ${\cal F}$
will diverge at those loci in the non-perturbative moduli space
where BPS states become massless. These are the loci
of massless magnetic monopoles and
massless dyons plus other singular lines at strong coupling.
Using the string-string duality between the $N=2$ heterotic and type IIA/B
strings, the non-perturbative heterotic free energy is identical to the
classical free energy of the type II strings, where one sums over the
classical BPS spectrum. Thus  ${\cal F}$ is singular precisely
on the discriminant locus $\Delta$ of the (mirror) Calabi-Yau which,
for the particular IIB model with $h_{21}=3$ and
$h_{11}=243$ for example, is given in (\ref{disclocus}).
In the next chapter we will identify the $N=2$ BPS sum ${\cal F}$
with the
gravitational threshold function
on the heterotic side; on
the type II side this is given by the known  topological function
$F_1^{II}$ \cite{BerCecOogVa}.

\subsection{Perturbative and non-perturbative
$N=2$ gravitational threshold
corrections}

In $N=2$ supergravity a particular combination of higher derivative
curvature terms (namely of $C^2$ and ${\cal R} \tilde{\cal R}$) resides in the
square of the chiral Weyl superfield.  Its coupling to the abelian vector
multiplets is governed by a holomorphic function ${\cal F}_{grav}$.
Below, ${\cal F}_{grav}$ will be identified with the $N=2$ holomorphic
free energy ${\cal F}$.
We will, in the following, focus on the dependence of
${\cal F}_{grav}$ on $S,T$ and $U$.  The discussion given below
can, in principle, also be extended to the dependence of
${\cal F}_{grav}$ on additional Wilson line moduli.

In $N=2$ heterotic string compactifications one has at tree-level
that ${\cal F}_{grav} = 24 S$, where
$S=\frac{1}{g^2}-i \frac{\theta}{8 \pi^2}$.
The gravitational coupling $g^{-2}_{grav}$
is then given by $g^{-2}_{grav} = \Re {\cal F}_{grav} =
24 \Re S$.  At the 1-loop level, on
the other hand, ${\cal F}_{grav}$ reads
\cite{CaLuMo,KaLou,AGNTT,Cu}
\beqa
{\cal F}_{grav} = 24 S_{inv} + \frac{b_{grav}}{8 \pi^2} \log \eta^{-2}(T)
\eta^{-2}(U) + \frac{2}{4 \pi^2} \log(j(T)-j(U))
\label{fgrav}
\eeqa
where $b_{grav} = 46 + 2(n_H - n_V) = 48 - \chi, \chi=2(n_V -
(n_H-1))$.\footnote{$\chi$ is the Euler number of the associated
CY manifold in the dual Type IIA formulation of the theory (assuming
that there exists such a dual formulation).}
$n_V$ denotes the number of massless vector multiplets (not including
the graviphoton) and $n_H$ the number of massless hyper multiplets
in the $N=2$ heterotic string compactification.
Here, $S_{inv}=S + \sigma(T,U)$
denotes the invariant dilaton field \cite{WitKapLouLu}.
It was shown in \cite{WitKapLouLu} that
$\sigma = -\frac{1}{2} \partial_T \partial_U h^{(1)} -
\frac{1}{8 \pi^2} \log (j(T) - j(U))$.
The term proportional
to $\log(j(T)-j(U))$ in (\ref{fgrav}) reflects the fact that
there are points of symmetry enhancement in the classical
$(T,U)$-moduli space
at which additional BPS states become massless \cite{CaLuMo}.
${\cal F}_{grav}$ has the correct modular weight
to render the
perturbative gravitational coupling $g^2_{grav}$  invariant
under the perturbative duality group
$SL(2,{\bf Z})_T\times SL(2,{\bf Z})_U\times {\bf Z}_2^{T
\leftrightarrow U}$
\beqa
\frac{1}{g^2_{grav}} = \Re {\cal F}_{grav} + \frac{b_{grav}}{16 \pi^2} (\log
\frac{M^2_{Planck}}{p^2} + K)
+ \frac{12(3-n_V)}{16 \pi^2} \log(S+\bar{S})
\label{grav}
\eeqa
where $K$ denotes the tree-level K\"{a}hler potential
$K = -\log (S+ {\bar S})(T + {\bar T})(U + {\bar U})$.
Note that there is an additional dependence on
$\log(S+\bar{S})$ in (\ref{grav}).\footnote{We thank Jan Louis for
pointing this out to us.}
The 1-loop corrected gravitational coupling (\ref{grav})
can also be written as follows
\beqa
\frac{1}{g^2_{grav}} &=& 12 \left( S + \bar{S} + V_{GS} \right) +
\frac{b_{grav}}{16 \pi^2} \log\frac{M^2_{string}}{p^2}
+ \frac{12(3-n_V)}{16 \pi^2} \log(S+\bar{S}) \nonumber\\
&+& \Delta_{grav}
\label{gravpl}
\eeqa
where
\beqa
\Delta_{grav} &=& 12(-V_{GS} + \sigma + \bar{\sigma})
+ \frac{b_{grav}}{16 \pi^2} \hat{K} \nonumber\\
&+& \Re \left(
\frac{b_{grav}}{8 \pi^2} \log \eta^{-2}(T)
\eta^{-2}(U) + \frac{2}{4 \pi^2} \log(j(T)-j(U)) \right)
\label{dgrav}
\eeqa
Here, $M^2_{Planck} \propto (S + \bar{S}) M^2_{string}$ and
$\hat{K} = - \log (T + {\bar T})(U + {\bar U})$.
$V_{GS}$ denotes the Green-Schwarz term and
$S + \bar{S} + V_{GS}$ denotes the true loop counting parameter
of the heterotic string.
Finally, note that (\ref{gravpl}) can also be written as
\beqa
\frac{1}{g^2_{grav}} &=&
\frac{b_{grav}}{16 \pi^2} \log\frac{M^2_{Planck}}{p^2} + \frac{1}{16 \pi^2}
F_1
\eeqa
where
\beqa
F_1&=&\log\lbrace\exp\lbrack({17\over 3}+{5\over 3}n_V+{1\over 3}n_H)K\rbrack
\det K_{i\bar{j}}^{-2}e^{
8 \pi^2 ({\cal F}_{grav}+\bar{\cal F}_{grav})} \rbrace
\eeqa
Here, $K_{i\bar{j}}$
denotes the tree-level K\"ahler metric
of the massless vector
multiplets.

As explained in section (4.1), the term proportional to
$\log \eta^{-2}(T)\eta^{-2}(U)$ arises from BPS states laying on
the orbit $m_1n_1 +  m_2 n_2 =0$, whereas the term proportional to
$\log(j(T)-j(U))$ arises from BPS states laying on the orbit
$m_1 n_1 + m_2 n_2 = 1$.  Thus, it is natural to conjecture
\cite{FerKouLuZwi,Vafa,HarvMoo} that ${\cal F}_{grav}$ is obtained
by summing over suitable orbits of BPS states, that is
\beqa
{\cal F}_{grav}
&\propto& {\cal F}=
(\sum_{M_I,N^I}^{{\rm vector}}-\sum_{M_I,N^I}^{{\rm hyper}})\log
  (M_IP^I+iN^IQ_I)
\label{gravthre}
\eeqa
Here, the period vector $(P^I,iQ_I)$ entering in (\ref{gravthre})
is given by the classical period vector (\ref{clperiod}).
Comparing (\ref{gravthre})
 with (\ref{fgrav}) shows that the tree-level piece ${\cal F}_{grav} =24 S$
should be due to BPS states as well, that is it should arise from
(\ref{gravthre}) when taking
$S \rightarrow \infty$.  For instance, it could arise from a term in
${\cal F}_{grav}$ of the type\footnote{Here, $\check{S} = 4 \pi S =
\frac{4 \pi}{g^2} - i \frac{\theta}{2 \pi} \label{stilde}$.  Then,
under the axionic
shift $\theta \rightarrow \theta + 2 \pi$, $\check{S} \rightarrow
\check{S} - i$.}
 $\log \eta^{-2}
(\check{S}) = \sum_{(s,p)\ne(0,0)} \log(s+ip\check{S})$ in the limit
$\check{S} \rightarrow \infty$.  Inspection of the mass formula
(\ref{massphya})
shows that such a term could indeed arise.

${\cal F}_{grav}$ will, in general, receive non-perturbative corrections.
It is natural to conjecture that the
non-perturbatively corrected ${\cal F}_{grav}$ will be given as in
(\ref{gravthre}), where this time the period vector $(P^I,iQ_I)$
is the non-perturbative period vector (\ref{npperiod}).
Finally note that, whereas on the heterotic side
$F_1$ describes the gravitational threshold
function, it is the known topological function $F_1^{II}$ on the
type II side \cite{BerCecOogVa}.

Consider the 1-loop corrected gravitational coupling (\ref{gravpl}).
In general, it is
difficult to compute
$\Delta_{grav}$ exactly at the 1-loop level.  For the s=0 model
(which has a  gauge group $G=E_8 \times E_7 \times U(1)^4$ at generic points
in the moduli space)
discussed recently in \cite{HarvMoo}, however, this can be done using
the technology introduced there, as follows.

It was shown in \cite{HarvMoo} that
the Green-Schwarz term $V_{GS}$ is given by
$V_{GS} = \frac{2}{16 \pi^2} \Delta_{univ}$ with $\Delta_{univ}$
given in equation (4.4) of \cite{HarvMoo}, that is\footnote{Here, $h^{(1)}$
denotes the 1-loop correction to the prepotential $F$
given in (\ref{npprep}), that is $h^{(1)} =
f^1$.}
\beqa
V_{GS} &=& \frac{2}{-(\Re y)^2}
\Re \left(h^{(1)} - y_1^a \partial_{y^a} h^{(1)}\right)
\nonumber\\
&=&
\frac{2(h^{(1)} + \bar{h}^{(1)}) - (y^a + \bar{y}^a)(\partial_{y^a}h^{(1)}
+ \partial_{\bar{y}^a} \bar{h}^{(1)})}{(T+\bar{T})(U+\bar{U})}
\eeqa
where
$y = (y_{+},y_{-})=(T,U),
y_1 = \Re y $.
It is convenient to introduce a
coupling $\tilde{S} = S - \frac{1}{2}
\partial_T \partial_U h^{(1)}$.  Then, it was shown in \cite{HarvMoo} that
$V_{GS}$ and $\tilde{S}$ satisfy the following
differential equation
\beqa
\frac{1}{2}\left(-V_{GS} + \tilde{S} - S + \tilde{\bar{S}} - \bar{S}\right)
&=& - \frac{1}{12} \frac{1}{16 \pi^2} \left( \tilde{I}_{2,2} - I_{2,2} \right)
+ \frac{1}{8 \pi^2} ( \log \Psi + \log \bar{\Psi}) \nonumber\\
&+&
\frac{b(E_8)}{16 \pi^2} \log(-y^2_1)
\label{diff}
\eeqa
Inserting (\ref{diff}) into $\Delta_{grav}$ in (\ref{dgrav}) gives that
\beqa
\Delta_{grav} &=& 24 \left(
- \frac{1}{12} \frac{1}{16 \pi^2} \left( \tilde{I}_{2,2} - I_{2,2} \right)
+ \frac{1}{8 \pi^2} ( \log \Psi + \log \bar{\Psi}) +
\frac{b(E_8)}{16 \pi^2} \log(-y^2_1) \right) \nonumber\\
&-& 12 \left( \frac{1}{8 \pi^2} \log(j(T)-j(U)) +
\frac{1}{8 \pi^2} \log(j(\bar{T})-j(\bar{U}))\right) \nonumber\\
&+& \frac{b_{grav}}{16 \pi^2} \hat{K}
+ \Re \left(
\frac{b_{grav}}{8 \pi^2} \log \eta^{-2}(T)
\eta^{-2}(U) + \frac{2}{4 \pi^2} \log(j(T)-j(U)) \right)
\label{sexp}
\eeqa
Using that \cite{HarvMoo}
\beqa
\log \Psi &=& \frac{1}{2} \log (j(T) - j(U)) + \frac{1}{2} b(E_8)
\log\eta^2(T) \eta^2(U) \nonumber\\
I_{2,2} &=& - c_3(0) \log(T + \bar{T})(U + \bar{U}) - 2 \log|j(T)-j(U)|^2
\nonumber\\
&-&  c_3(0) \log|\eta^2(T) \eta^2(U)|^2 + {\rm constant}
\label{exp}
\eeqa
and inserting (\ref{exp}) into (\ref{sexp})
gives that
\beqa
\Delta_{grav} = - \frac{2}{16 \pi^2} \tilde{I}_{2,2}
\label{smoore}
\eeqa
where we have used that
$b_{grav}= -2\tilde{c}_1(0) = 528, c_3(0) = -984 , b(E_8) = -60$.
$\tilde{I}_{2,2}$ is given by \cite{HarvMoo}
\beqa
\tilde I_{2,2}
=\int\limits_{\cal F} \frac{d^2\tau}{\tau_2}
         [Z_{2,2}\frac{E_4E_6}
{\eta^{24}}(E_2-\frac{3}{\pi\tau_2})-\tilde{c}_1(0)]
\eeqa
Using the results of \cite{HarvMoo}, it
is straightforward to show that $\tilde{I}_{2,2}$ is related
to the "new supersymmetric index" of \cite{CecFenIntVa} as follows
\beqa
\tilde{I}_{2,2}
=-\frac{1}{2}\int\limits_{\cal F} \frac{d^2\tau}{\tau_2}
    [-\frac{i}{\eta^2}Tr_R J_0(-1)^{J_0}q^{L_0-22/24}\bar{q}^{\tilde{L}_0-9/24}
    (E_2-\frac{3}{\pi\tau_2})-b_{grav}]
\label{ind}
\eeqa
Expression (\ref{ind}), on the other hand, was shown in \cite{HarvMoo}
to be due to BPS states only.  That is, $\Delta_{grav}$ is indeed
due to BPS states, only.  The integral $\tilde{I}_{2,2}$ was
explicitly evaluated in \cite{HarvMoo} and is given by
\beqa
\tilde{I}_{2,2} &=& 4 \Re \left( \sum_{r>0} [ \tilde{c}_1 ( - \frac{r^2}{2})
Li_1(e^{-2 \pi r \cdot y}) + \frac{6}{\pi y^2_1} c_1(-\frac{r^2}{2})
{\cal P}(r \cdot y) ] \right) \nonumber\\
&+& \tilde{c}_1(0) \left( - \log[-y^2_1] - \kappa \right) +
\frac{1}{y^2_1}[\tilde{d}^{2,2}_{abc}y^a_1 y^b_1 y^c_1 + \delta]
\label{tili}
\eeqa
Note that $\tilde {I}_{2,2}$ possesses a $T-U$ chamber dependence, i.e.
$\tilde I_{2,2}(\Re T>\Re U)\neq \tilde I_{2,2}(\Re U>\Re T)$.
The exact expression for the
1-loop corrected gravitational coupling (\ref{gravpl}) then follows
from (\ref{tili}) and from the
explicit evaluation of $V_{GS}$ given in \cite{HarvMoo}
\footnote{Note that the polynomial term is missing in equation (4.28)
of \cite{HarvMoo}.}
\beqa
V_{GS} = \frac{1}{y^2_1} \frac{2}{(2 \pi)^3} \Re \left(\sum_{r>0}
c_1(-\frac{r^2}{2}) {\cal P}(r \cdot y) \right) + \frac{1}{96 \pi^2}
\frac{1}{y^2_1} \left(\tilde{d}^{2,2}_{abc}y^a_1 y^b_1 y^c_1 + \delta \right)
\eeqa
Similarly, the exact expression for the
1-loop corrected holomorphic coupling ${\cal F}_{grav}$
(\ref{fgrav}) is given by
\beqa
{\cal F}_{grav} &=& 24 \left( S -
\frac{1}{768 \pi^2} \partial_T \partial_U
\left( \tilde{d}^{2,2}_{abc} y^a y^b y^c \right)
- \frac{1}{8 \pi^2} \log (j(T) - j(U)) \right. \nonumber\\
&+& \left. \frac{1}{8 \pi^2}
\sum_{r > 0} c_1(kl) kl Li_{1}(e^{-2\pi r
 \cdot y})  \right)
+ \frac{b_{grav}}{8 \pi^2} \log \eta^{-2}(T)
\eta^{-2}(U) \nonumber\\
&+& \frac{2}{4 \pi^2} \log(j(T)-j(U))
\label{fexp}
\eeqa
Now, consider taking the limit $T \rightarrow \infty$ of (\ref{fexp})
(keeping $U$ finite).
Then, using that $\log \eta^{-2}(T)
\rightarrow \frac{\pi}{6}T, \log j(T) \rightarrow 2 \pi T$,
it is straightforward to
show that
\beqa
{\cal F}_{grav} \rightarrow  24 S
\label{flim}
\eeqa
Note that a possible linear $T$-dependence drops out in this limit!
(\ref{flim}) is nothing but the holomorphic gravitational coupling
of an $N=4$ heterotic compactification.  Thus, it is suggestive to
interprete the limit
$\Re S > \Re T \rightarrow \infty$ in this model
as a limit in which one obtains
an $N=4$ like situation.

$N=4$
string/string/string triality, on the other hand,
says that $N=4$ compactifications of the heterotic string and of the
type II string are related through exchange symmetries $S \leftrightarrow T$
or $S \leftrightarrow U$
\cite{DuffLiuRahm}.  Thus $N=4$ string/string/string triality
together with our discussion in section (3.3) then
suggests that there might be an $S \leftrightarrow T,U$ exchange
symmetry at the non-perturbative level in the $s=0$ model of \cite{HarvMoo}!
As discussed in section (3.3), such an exchange symmetry
is made possible due to those short $N=2$ BPS multiplets which from
an $N=4$ point of view are intermediate BPS states.
The direct evalution of
non-perturbative corrections to ${\cal F}_{grav}$ is very hard,
because in order to evaluate
${\cal F}_{grav} \propto
(\sum_{M_I,N^I}^{{\rm vector}}-\sum_{M_I,N^I}^{{\rm hyper}})
\log (M_IP^I+iN^IQ_I)$ knowledge of the non-perturbative period vector
$\Omega$ is needed. The existence of an exchange symmetry
$S \leftrightarrow T$, on the other hand, would allow one
to produce quantitative statements about ${\cal F}_{grav}$ in
a certain strong coupling regime.

How would such an exchange symmetry
act on ${\cal F}_{grav}$?  Consider first the
2 parameter model
$P_{1,1,2,2,6}(12)$ of \cite{CanOsFoKaMo}. It was observed in
\cite{KLM} that this model indeed possesses
an exchange symmetry $S \leftrightarrow T$
at the non-perturbative
level.  In this model,
the holomorphic gravitational coupling ${\cal F}^{top}_1
=  \frac{2  \pi^2 }{3}
{\cal F}_{grav}$
enjoys an instanton
expansion of the following type
\beqa
{\cal F}^{top}_1 = - \frac{2 \pi i}{12} c_2 \cdot (B+ iJ)
-\sum_{j,k \ge 0} \left( 2 d_{jk} \log \eta (q_1^j q_2^k) + \frac{1}{6} n_{jk}
\log(1-q_1^j q_2^k) \right)
\eeqa
where \cite{CanOsFoKaMo,KLM}
$q_i = e^{2 \pi i t_i}$, $t_1 = iT$, $t_2 = i(\check{S}-T)$, $q_1^j q_2^k =
e^{-2 \pi  T (j-k)} e^{-2 \pi  k \check{S}}$
and where $c_2 \cdot (B + iJ) = 24 t_2 + 52 t_1 = i(24 \check{S} + 28 T)$.
Here,
$\check{S} = 4 \pi S$ denotes the redefined dilaton which enjoys
modular properties (see footnote \ref{stilde}).
In the weak coupling limit $\Re T < \Re S \rightarrow \infty$,
$q_1^j q_2^k \rightarrow 0$,
and so
\beqa
{\cal F}^{top}_1 \rightarrow  - \frac{2 \pi i}{12} c_2 \cdot (B+ iJ)
=  \frac{2 \pi }{12} \left( 24 \check{S} + 28 T \right)
\label{sttop}
\eeqa
This expression agrees with the large $T$ limit of the
known 1-loop expression \cite{KaLou}
\beqa
{\cal F}_{grav}=24 S_{\rm inv}
+{1\over 4\pi^2}\log(j(T)-j(1))-{300\over 4\pi^2}\log
\eta^2(T)\label{ssss}
\eeqa
provided that $S_{\rm inv}=S+\sigma(T)$ goes in the limit $T\rightarrow
\infty$ as $S_{\rm inv}\rightarrow S-{1\over 4\pi}T$.

In the strong coupling
limit $\Re T > \Re S\rightarrow \infty$, on the other hand,
one has that $q_1^j q_2^k$ is only non-vanishing for $k >j$.
For $k>j$ one has that $d_{jk}=0$, $n_{jk} = 2 \delta_{j0} \delta_{k1}$.
It follows that
\beqa
\sum_{j,k\ge 0} n_{jk}
\log(1-q_1^j q_2^k)  = 2 \log(1-q_2) = 2 \log q_2 = 4 i \pi t_2
= -4  \pi (\check{S}-T)
\eeqa
Then
\beqa
{\cal F}^{top}_1 \rightarrow  - \frac{2 \pi i}{12} c_2 \cdot (B+ iJ)
+ \frac{4  \pi  }{6} (S-T)
= \frac{2 \pi }{12} \left( 28 \check{S} + 24 T \right)
\label{ts}
\eeqa
Combining (\ref{sttop}) and (\ref{ts}) gives that
as $S \rightarrow \infty , T \rightarrow \infty$
\beqa
{\cal F}^{top}_1 =
 \frac{2 \pi }{12} \left( (24 \check{S} + 28 T ) \theta (\check{S}-T)
+ ( 28 \check{S} + 24 T) \theta (T - \check{S}) \right)
\label{ftoplim}
\eeqa
which exhibits the exchange symmetry $\check{S} \leftrightarrow T$.
\footnote{Taking $T \rightarrow \infty$ corresponds to
decompactification \cite{AntFerTay} to 5 dimensions.  In 5 dimensions
there is a discontinuity at $t=1$ (where $t$ is the 5D modulus)
corresponding to the non-perturbative singularity at $\check{S}=T$ in 4
dimensions
\cite{AntFerTay}.}  Note that (\ref{ftoplim}) exhibits an $\check{S}-T$
chamber dependence. Applying
the $\check{S} \leftrightarrow T$ exchange symmetry on the
1-loop expression eq.(\ref{ssss}) we obtain for large $T$
but arbitrary $S$ the non-perturbative
gravitational coupling as
\beqa
{\cal F}_{grav}=24 T_{\rm inv}
+{1\over 4\pi^2}\log(j(\check{S})-j(1))-{300\over 4\pi^2}\log
\eta^2(\check{S})
\eeqa

Let us then assume that the $s=0$ model of \cite{HarvMoo}
also possesses an exchange symmetry $\check{S} \leftrightarrow T$
(and similarly for $\check{S} \leftrightarrow U$).
Then,
in view
of (\ref{fexp}),
the non-perturbative ${\cal F}_{grav}$ should in the limit $T \rightarrow
\infty$
be given to all orders in
$\check{S}$ and $U$ by
\beqa
{\cal F}_{grav} &=& 24 \left( \frac{T}{4 \pi} -
\frac{1}{768 \pi^2} \partial_{\check{S}} \partial_U
\left( \tilde{d}^{2,2}_{abc} \tilde{y}^a \tilde{y}^b \tilde{y}^c \right)
- \frac{1}{8 \pi^2} \log (j(\check{S}) - j(U)) \right. \nonumber\\
&+& \left. \frac{1}{8 \pi^2}
\sum_{r > 0} c_1(kl) kl L_{i_1}(e^{-2\pi r
 \cdot \tilde{y}})  \right)
+ \frac{b_{grav}}{8 \pi^2} \log \eta^{-2}(\check{S})
\eta^{-2}(U) \nonumber\\
&+& \frac{2}{4 \pi^2} \log(j(\check{S})-j(U))
\label{ftexa}
\eeqa
where $\tilde{y}=(\check{S},U)$. The logarithmic singularity at $\check{S}=U$
corresponds to the $SU(2)$ gauge symmetry enhancement along this line,
which is further enhanced to $SU(2)^2$, $SU(3)$ at $\check{S}=U=1$ and
$\check{S}=U^{-1}=\rho$ respectively.
It follows that $g_{\rm grav}^2$ is invariant under $SL(2,{\bf Z})_{\check{S}}
\times
SL(2,{\bf Z})_U\times {\bf Z}_2^{\check{S}\leftrightarrow U}$ for large $T$.
Taking the limit $\check{S} \rightarrow \infty$ of (\ref{ftexa})
yields that
\beqa
 {\cal F}_{grav} \rightarrow \frac{6}{\pi} T
\label{tlim}
\eeqa
(\ref{tlim}) gives the holomorphic gravitational coupling
for an $N=4$ compactification of the type IIA string.
Combining both (\ref{flim}) and (\ref{tlim}) yields that
\beqa
 {\cal F}_{grav} = \frac{6}{\pi}\left ( T \theta ( T - \check{S} ) +
\check{S} \theta ( \check{S}-T) \right)\label{fivedim}
\eeqa
in analogy to (\ref{ftoplim}).

Our results contain, as a subcase, the five dimensional results
of \cite{AntFerTay} which discussed the
behavior of the model in the two limits $\Re S>\Re T\rightarrow
\infty$ and $\Re T>\Re S\rightarrow\infty$.
In five dimensions the
$\theta$-function
discontinuities in eq.(\ref{fivedim})
are again due to non-perturbative states becoming
massless at $\check{S}=T$ and $\check{S}=U$
\cite{AntFerTay}. However the $\check{S}\leftrightarrow  T$ exchange symmetry
provides also information in the entire strong coupling region
$T\rightarrow\infty$ for arbitrary $\check{S}$ (and $U$.)
In particular this symmetry predicts the further gauge symmetry
enhancement at the strong coupling points $\check{S}=U=1$
and $\check{S}=U^{-1}=\rho$.

\section{Conclusions}

In this paper, we
studied the BPS spectrum in $D=4, N=4$ heterotic string compactifications.
These BPS states can either fall into short or into intermediate
multiplets.  As pointed out in \cite{DuffLiuRahm},
the string/string/string triality
conjecture between $N=4$ compactifications of the heterotic, the
type IIA and the type IIB string
implies, for instance, that the BPS spectrum
of the heterotic and of the type IIA
string are mapped into each other under the exchange $S \leftrightarrow T$.
The BPS mass spectrum of the heterotic (type IIA) string is, however,
not symmetric under this exchange of $S$ and $T$.  This is due to
the fact that BPS masses in $D=4, N=4$ compactifications are
given by the maximum of the 2 central charges $|Z_1|^2$ and
$|Z_2|^2$.  On the other hand,
states, which from the $N=4$ point of view are intermediate,
are
actually short from the $N=2$ point of view.  This then leads
to the possibility that the BPS spectrum of certain
$N=2$ heterotic compactifications is actually symmetric under
the exchange of $S$ and $T$.  Since contributions to the
holomorphic gravitational
coupling ${\cal F}_{grav}$ arise from BPS states only
(as shown in \cite{HarvMoo}
for 1-loop contributions), it follows
that ${\cal F}_{grav}$ should exhibit a symmetry under the
exchange of $S$ and $T$.
As an example of an $N=2$ compactification
we took the $s=0$ model of \cite{HarvMoo} and computed
the exact 1-loop contribution to the holomorphic gravitational
coupling ${\cal F}_{grav}$ using the technology introduced in
\cite{HarvMoo}.  We then showed that in the decompactification
limit $T \rightarrow
\infty$ at weak coupling one recovers the tree level holomorphic gravitational
coupling.  This $N=4$ like situation then suggests that
the $N=4$ triality exchange symmetries are actually realised
as exchange symmetries $S\leftrightarrow T$ and $S \leftrightarrow U$
in the
$s=0$ $N=2$ heterotic
model.  Assuming that there are indeed
such exchange symmetries in the $s=0$ model allows one
to evaluate non-perturbative corrections to the
gravitational couplings in some of the non-perturbative regions (chambers)
in this particular heterotic model.

\section{Acknowledgement}

We would like to thank L. Ib\'a\~nez and F. Quevedo for their participation
at the initial stages of this work.
We also would like to thank  L. Alvarez-Gaum\'e, J. Buchbinder,
A. Chamseddine, M. Cveti\v{c},
E. Derrick,  S. Ferrara, R. Khuri, A. Klemm, W. Lerche, J. Louis, P. Mayr and
A. Sen for fruitful discussions.
Two of us (D.L. \& S.-J.R.) are grateful to the Aspen Center of Physics, where
part of this work was performed.  The work of T.M. is supported by DFG.
The work of S.-J.R. is supported by U.S.NSF-KOSEF Bilateral Grant, KRF 
Nondirected Research Grant 81500-1341, KOSEF Purpose-Oriented Research
Grant 94-1400-04-01-3, KRF International Collaboration Grant, Ministry of
Education BSRI-94-2418 and KOSEF-SRC Program.

\section{Appendix A}

\setcounter{equation}{0}

 \hspace*{.3in}
We will, in this appendix, discuss
orbits and invariants of duality groups.
For many purposes, like the computation of the BPS sums in the previous
sections, it is very useful to consider subsets of BPS states which fall
into socalled orbits of the duality group.
An orbit of a group on a set is a subset that is invariant under
the group action. To divide a set into orbits one must therefore
take group invariant constraints. We are interested in finding
orbits of the group
\be
SL(2,{\bf Z})_S\otimes  SL(2, {\bf Z})_T \otimes SL(2, {\bf Z})_U \otimes
{\bf Z}_2^{(\rm Mirror)}
\label{HetTDualGroup}
\eq
and some of its subgroups
on the set of (short or intermediate) BPS states.
Here ${\bf Z}_2^{(\rm Mirror)}$ denotes the perturbative
${\bf Z}_2$ group which permutes the two moduli of the theory
under consideration, i.e. $T \leftrightarrow U$ for the heterotic
theory. In the following we will for definiteness always deal
with the  $N=4$ heterotic string, if not specified otherwise.

As
pointed out in eq.(\ref{tdual}) and below, the quantities
\be
\left( \begin{array}{c} \hat{M}_2 \\ \hat{M}_0 \\ \end{array} \right),\;\;\;
\left( \begin{array}{c} \hat{M}_1 \\ \hat{M}_3 \\ \end{array} \right),\;\;\;
\left( \begin{array}{c} \hat{N}_2 \\ \hat{N}_0 \\ \end{array} \right),\;\;\;
\left( \begin{array}{c} \hat{N}_1 \\ \hat{N}_3 \\ \end{array} \right)
\label{ttdual}
\eq
transform as $SL(2,{\bf Z})_T$ vectors, i.e. by multiplication
with $\left( \begin{array}{cc} a&c\\b&d\\ \end{array} \right)$.
Analogously, the quantities
\be
\left( \begin{array}{c} \hat{M}_3 \\ \hat{M}_0 \\ \end{array} \right),\;\;\;
\left( \begin{array}{c} \hat{M}_1 \\ \hat{M}_2 \\ \end{array} \right),\;\;\;
\left( \begin{array}{c} \hat{N}_3 \\ \hat{N}_0 \\ \end{array} \right),\;\;\;
\left( \begin{array}{c} \hat{N}_1 \\ \hat{N}_2 \\ \end{array} \right)
\label{uudual}
\eq
transform as $SL(2,{\bf Z})_U$ vectors.
The non--perturbative $SL(2,{\bf Z})_S$
acts by
\be
M_S = \left( \begin{array}{cc}
d \cdot {\bf 1}_4 & b \cdot {\bf 1}_4 \\
c \cdot {\bf 1}_4 & a \cdot {\bf 1}_4 \\
\end{array} \right),
\eq
where the representation space is now spanned by the vectors
${\bf V}=(\hat{M}_0,\ldots,\hat{N}_3)$ consisting of all electric and magnetic
quantum numbers. In other words $(\hat{M}_I, \hat{N}_I)$ transforms as a
$SL(2,{\bf Z})_S$ vector for fixed $I$.

Let us first discuss orbits and invariants of a single $SL(2,{\bf Z})$,
which for definiteness we take to be $SL(2,{\bf Z})_S$. The eight--dimensional
representation on the quantum numbers is of course reducible and decomposes
into the four irreducible two--dimensional representations specified above.
We begin by looking at orbits and invariants associated to an irreducible
two--dimensional representation. In order to characterize orbits, we would
like to construct invariants out of vectors ${\bf v}$,
which could then lable the
orbits.
As is well known the only invariant
tensor of the corresponding continuous group $SL(2,{\bf R})$ is the
$\epsilon$ tensor and the related invariant is nothing but the
antisymmetric scalar product
\be
({\bf v}, {\bf w}) = \epsilon^{ij} v_i w_j
\eq
Due to antisymmetry we cannot construct a non--trivial invariant out of a
single vector, since $({\bf v}, {\bf v})= 0$. How then characterize
orbits? First note that the vector $(1,0)^T$ can be mapped to any other
vector ${\bf v} \not= 0$ by an $SL(2,{\bf R})$ transformation. Therefore
the continuous group has precisely two orbits, namely the zero vector
$\{ {\bf v}=0 \}$ and the punctured plane $\{ {\bf v} \not= 0 \}$.
The non--existence of a non--trivial invariant associated to a single vector
reflects the fact that all vectors ${\bf v} \not=0$ are related by group
transformation. Conversely groups like $SO(2)$ where one has such
invariants (the length) have orbits that are labled by the invariant
(circles of a given radius).

Clearly the orbit $\{ {\bf v} \not= 0 \}$ becomes highly reducible,
when switching to the discrete group $SL(2,{\bf Z})$. To see this
just note that $(p,0)^T$ and $(q,0)^T$, $p,q \in {\bf Z}$ cannot
be related by a $SL(2,{\bf Z})$ for coprime $p,q$. However the
discrete version of ${\bf v} \not=0$, namely
$\{ (p,q) \not= (0,0) | p,q \in {\bf Z} \}$ is precisely the kind of orbit
that one needs, since various modular forms including all Eisenstein
series and (using $\zeta$ regularization) the Dedeking $\eta$ function
can be expressed as sums over a two dimensional lattice with the origin
excluded.

Let us next discuss the reducible representation of $SL(2,{\bf Z})_S$
on the eight electric and magnetic quantum numbers $\hat{M}_I, \hat{N}_I$.
Since this decomposes into four irreducible representations
we can now construct
six non--trivial (this means generically non--vanishing) invariants
by taking mutual scalar products between the various irreducible
parts. These invariants $\hat{M}_I \hat{N}_J - \hat{N}_I \hat{M}_J$,
$I < J$ can be arranged into an antisymmetric invariant matrix:
\be
\hat{M}_I \hat{N}_J - \hat{M}_J \hat{N}_I =: C_{IJ}\label{cijorbit}
\eq
Note that this matrix is in fact the exterior product of the
electric and magnetic part of the vector of quantum numbers:
\be
\hat{\bf M} \wedge \hat{\bf N} = {\bf C}
\eq
Further note that this product vanishes if and only if the electric and
magnetic part are parallel:
\be
\hat{\bf M} \wedge \hat{\bf N} = 0
\Longleftrightarrow \vec P^{\rm het} || \vec Q^{\rm het}
\eq

The groups $SL(2,{\bf Z})_T$ and $SL(2,{\bf Z})_U$ can be
treated in a similar way. The simplest way to obtain the corresponding
invariants is to apply the duality transformations $S \leftrightarrow T$
and $S \leftrightarrow U$, respectively. Obviously the
six non--trivial invariants of $SL(2,{\bf Z})_T$ ($SL(2,{\bf Z})_U$)
vanish simultanously if and only if electric and magnetic vector
of the IIA (IIB) theory are parallel.

Let us now consider orbits and invariants for products of two
$SL(2,{\bf Z})$ groups. For defineteness we will take the
T duality group $SL(2,{\bf Z})_T \otimes SL(2,{\bf Z})_U$ of
the heterotic string. The eight--dimensional representation space
splits under this group into two irreducible four--dimensional
representations spanned by the electric and magnetic parts.
The invariant tensor $\epsilon \otimes \epsilon$ is represented
by a symmetric matrix on each irreducible part which is easily
found to be conjugated to the standard $SO(2,2)$ invariant metric,
as expected from the local isomorphism
$SL(2,{\bf R}) \otimes SL(2,{\bf R}) \simeq SO(2,2)$.
The corresponding invariant is the $SO(2,2)$ scalarproduct $\la , \ra$,
which
reads in our parametrization:
\be
\la{\bf V}, {\bf W} \ra = V_0 W_1 + V_1 W_0 - V_3 W_2 - V_2 W_3
\eq
Therefore one can construct three non--trivial invariants out of
the quantum numbers $(\hat{\bf M}, \hat{\bf N})^T$ namely the scalar
products $\la \hat{\bf M}, \hat{\bf M} \ra$,
$\la \hat{\bf N}, \hat{\bf N} \ra$
and $\la \hat{\bf M}, \hat{\bf N} \ra$. $SO(2,2)$ orbits of the form
$\la \hat{\bf M}, \hat{\bf M}\ra = const$ play an important role
in perturbative
threshold corrections and they are indeed related to $SO(2,2,{\bf Z})$ modular
forms. Orbits of the form $\la \hat{\bf M}, \hat{\bf N} \ra = const$
also play some role because they appear as suborbits of
$SL(2,{\bf Z})_S \otimes SL(2,{\bf Z})_T \otimes SL(2,{\bf Z})_U$
orbits. Finally note that the $SO(2,2)$ scalar product is also
manifestly invariant under the perturbative heterotic mirror
symmetry ${\bf Z}_2^{(T \leftrightarrow U)}$ and therefore it is
invariant under the full perturbative heterotic duality group
$SL(2,{\bf Z})_T \otimes
SL(2,{\bf Z})_U \otimes {\bf Z}_2^{(T \leftrightarrow  U)}$.

Let us now discuss orbits and invariants of the full
duality group
$SL(2,{\bf Z})_S \otimes SL(2,{\bf Z})_T
\otimes SL(2,{\bf Z})_U \times {\bf Z}_2^{(T \leftrightarrow U)}$.
Under this group our eight--dimensional representation is
irreducible and since the invariant tensor
$\epsilon \otimes \epsilon \otimes \epsilon$ is antisymmetric we cannot
construct an invariant out of the vector $(\hat{\bf M}, \hat{\bf N})^T$.
This situation is similar to that of the irreducible two--dimensional
representation of a single $SL(2,{\bf Z})$. However the non--existence
of an invariant {\bf number} that one can assign to an orbit does not
mean that there are no invariant {\bf equations} that characterize orbits.
A closer inspection shows that the six $SL(2,{\bf Z})_S$ invariants
$C_{IJ}$ are not invariant under
$SL(2,{\bf Z})_T \otimes SL(2,{\bf Z})_U
\otimes {\bf Z}_2^{(T \leftrightarrow U)} $
for generic values, but that they are invariant if and only if
they vanish. Thus either $C_{IJ} = 0$ or $C_{IJ} \not= 0$ are invariant
equations which decompose the representation space into disjoint orbits.
These conditions are the analogue of ${\bf v} = 0$, ${\bf v} \not= 0$
in the case of a single $SL(2,{\bf Z})$.
In geometrical terms one can say that although the 'angle' between
the electric and magnetic part is not preserved, parallelity
is respected.

Let us therefore summarize that the condition
\be
\hat{\bf M} \wedge \hat{\bf N} = 0
\eq
defines an orbit of the full group
$SL(2,{\bf Z})_S \otimes SL(2,{\bf Z})_T \otimes SL(2,{\bf Z})_U \otimes
{\bf Z}_2^{(T \leftrightarrow U)}$,
which is singled out by (i) the simultanous vanishing of all
$SL(2,{\bf Z})_S$ invariants and (ii) by the parallel alignement of
the electric and magnetic quantum numbers.
This orbit, which we
call the $S$ orbit, is clearly a special, non--generic subset of vectors.
Note that it contains the short $N=4$ BPS multiplets of the
heterotic theory. Note also that $\hat{\bf M}$ and $\hat{\bf N}$ being
parallel implies that the quantum numbers are pairwise proportional, that is
$s \hat{M}_I = p \hat{N}_I$, $(\exists p,s \in {\bf Z})$.

Obviously we can construct two further distinguished orbits of
$SL(2,{\bf Z})_S \otimes SL(2,{\bf Z})_T \otimes SL(2,{\bf Z})_U$\footnote{
The ${\bf Z}_2^{(\rm Mirror)}$ symmetries will be discussed below.}
by applying the transformations $S \leftrightarrow T$ and
$S \leftrightarrow U$ to the $S$ orbit. The resulting orbits
will be called the $T$ and the $U$ orbit. They are singled out by
the simultanous vanishing of the six $SL(2,{\bf Z})_T$
($SL(2,{\bf Z})_U$) invariants and by parallel alignement of
the electric and magnetic quantum numbers of the IIA (IIB)
theory. Denoting these electric and magnetic quantum numbers
by
\be
\hat{\bf M}^{(A)} = (\hat{M}_0, \hat{N}_3, \hat{N}_0, \hat{M}_3)^T,
\;
\hat{\bf N}^{(A)} = (\hat{M}_2, \hat{N}_1, \hat{N}_2, \hat{M}_1)^T,
\eq
\be
\hat{\bf M}^{(B)} = (\hat{M}_0, \hat{N}_2, \hat{M}_2, \hat{N}_0)^T,
\;
\hat{\bf N}^{(B)} = (\hat{M}_3, \hat{N}_1, \hat{M}_1, \hat{N}_3)^T,
\eq
the T and the U orbit are characterized by
\be
\hat{\bf M}^{(A)} \wedge \hat{\bf N}^{(A)} = 0,\;\;\;
\hat{\bf M}^{(B)} \wedge \hat{\bf N}^{(B)} = 0,
\eq
respectively. The perturbative mirror symmetry
${\bf Z}_2^{(T \leftrightarrow U)}$ of the heterotic string
is mapped to the corresponding perturbative mirror symmetries
${\bf Z}_2^{(S \leftrightarrow U)}$ (
${\bf Z}_2^{(S \leftrightarrow T)}$) of the IIA (IIB) theory
by the non--perturbative duality transformations
$S \leftrightarrow T$ and $S \leftrightarrow U$. Thus the
$T$ orbit ($U$ orbit) is invariant under the full duality
group
$SL(2,{\bf Z})_S \otimes SL(2,{\bf Z})_T
\otimes SL(2,{\bf Z})_U \otimes {\bf Z}_2^{(S \leftrightarrow U)}$
($SL(2,{\bf Z})_S \otimes SL(2,{\bf Z})_T
\otimes SL(2,{\bf Z})_U
\otimes {\bf Z}_2^{(S \leftrightarrow T)}$) of the IIA (IIB) string.

Another even more special orbit is given by the constraint that
the non--trivial invariants of all three $SL(2,{\bf Z})$ groups
vanish simultanously. This gives the intersection of
the $S,T,U$ orbits and will therefore be called the $STU$ orbit.
The states in it fulfill
\be
\hat{M}_0 \hat{M}_1 = \hat{M}_2 \hat{M}_3
\label{STUorbit}
\eq
on top of $\hat{M}_I$, $\hat{N}_I$ being proportional.

One could also try to define orbits by setting the invariants
of only two $SL(2,{\bf Z})$ subgroups simultanously to zero.
But it turns out that then the invariants of the third
$SL(2,{\bf Z})$ are automatically also zero and we are back at the
$STU$ orbit.

Finally note that the 0 vector is trivially an invariant
suborbit of the $STU$ orbit, and that the $STU$ orbit is itself
an invariant suborbit of the $S$,$T$ and $U$ orbit. Therefore
disjoint invariant orbits are given bei 0, $STU$ - $0$,
$S$ - $STU$, $T$ - $STU$, $U$ - $STU$.

\section{Appendix B}

\setcounter{equation}{0}

 \hspace*{.3in}

In this appendix we investigate at which points in the $N=4$ heterotic
moduli space one can obtain, at least in principle, massless intermediate
spin 3/2 BPS states.
Clearly, at the special points of massless intermediate states
one has that
$|Z_1|^2=0$ and $\Delta Z^2=0$. In case that $S+\bar S\neq0$
this further implies that
\beqa
\hat M_0 -\hat M_1 TU + i\hat M_2 T + i\hat M_3 U &=& 0, \nonumber\\
\hat N_0  - \hat N_1TU
+ i \hat N_2 T + i \hat N_3 U &=& 0.
\label{masslessint}
\eeqa
Thus intermediate multiplets may become massless at special lines/points
in the $T,U$ moduli space for generic values of $S$.
First consider the line $T=U$.  It follows from (\ref{masslessint})
that the only states becoming massless at this line are the states
having $\hat M_2=-\hat M_3$, $\hat N_2 =- \hat N_3 =0$,
$\hat M_0=\hat M_1=\hat N_0=\hat N_1=0$. However,
for these states $\hat{M} \propto \hat{N}$, so that these
states are actually short, and not intermediate.

Next consider intermediate states becoming massless at the point $T=U=1$.
These are the states for which ${ \cal M}_{1,2} = 0$ at $T=U=1$:
\beqa
\hat{M}_0 -\hat{M}_1  + i (\hat{M}_2 + \hat{M}_3 ) = 0 \nonumber\\
\hat{N}_0  - \hat{N}_1
+ i (\hat{N}_2  +  \hat{N}_3 ) = 0
\label{massless}
\eeqa
Then, the only intermediate
states ${\bf V}=(\hat{M}_0,\dots,\hat{N}_3)$
satisfying (\ref{massless}) are as follows (we rescrict the
non-vanishing
charges to be $\pm 1$):
\beqa
a)~ \hat{M}_2 &=& - \hat{M}_3 =\pm 1, \hat{N}_0 = \hat{N}_1 =\pm 1\nonumber\\
b)~ \hat{M}_2 &=& - \hat{M}_3 =\pm 1, \hat{N}_0 = \hat{N}_1=\pm 1,
\hat{N}_2 = - \hat{N}_3 =\pm 1\nonumber\\
c) ~\hat{M}_0 &=&  \hat{M}_1= \pm 1, \hat{N}_2 = -\hat{N}_3=\pm 1 \nonumber\\
d) ~\hat{M}_0 &=&  \hat{M}_1=\pm 1 , \hat{M}_2 = - \hat{M}_3 =\pm 1,
\hat{N}_2 = -\hat{N}_3 =\pm 1\nonumber\\
e) ~\hat{M}_0 &=&  \hat{M}_1 =\pm 1, \hat{M}_2 = - \hat{M}_3=\pm 1 ,
\hat{N}_0 =  \hat{N}_1=\pm 1 \nonumber\\
f)~ \hat M_0 &=& \hat M_1=\pm 1,\hat N_0=\hat N_1=\pm 1
,\hat N_2=-\hat N_3=\pm 1\nonumber\\
g)~ \hat{M}_0 &=&   \hat{M}_1 =\pm 1, \hat{M}_2 = -\hat{M}_3 =\pm 1,
\hat{N}_0 =  \hat{N}_1 =\mp 1, \hat{N}_2 = - \hat{N}_3=\pm 1
\label{states}
\eeqa

Next consider the point $T=\bar U=\rho$.
Here dyons become massless  with the
following electric magnetic charge vectors
\beqa
\pm (1,1,0,0;1,0,1,-1)&,&\quad
\pm (1,1,0,0;-1,0,-1,1),\quad
\pm (1,1,0,0;0,-1,1,-1), \nonumber\\
\pm (1,1,0,0;0,1,-1,1)&,& \quad \pm (1,0,1,-1;1,1,0,0),\quad
\pm (1,0,1,-1;-1,-1,0,0), \nonumber\\
\pm (1,0,1,-1;0,-1,1,-1)&,&\quad \pm (1,0,1,-1;0,1,-1,1),\quad
\pm (0,1,-1,1;1,1,0,0), \nonumber\\
\pm (0,1,-1,1;-1,-1,0,0)&,& \quad \pm (0,1,-1,1;-1,0,-1,1),\nonumber\\
\pm (0,1,-1,1;1,0,1,-1)&.&\label{statesr}
\eeqa

Next, let us discuss the possible appearance
of massless intermediate multiplets for the
case of strong couplings, i.e. $S_1={\rm Re}S=0$. (Of course, via $S$-duality
one could equivalently consider weak coupling.)
Then one gets massless intermediate states if the following condition is
satisfied ($S_2={\rm Im}S$)
\beqa
\hat M_0 -\hat M_1 TU + i\hat M_2 T + i\hat M_3 U =
S_2(\hat N_0  - \hat N_1TU
+ i \hat N_2 T + i \hat N_3 U ).
\label{masslessinta}
\eeqa
Now consider one of the intermediate states given in (\ref{states}),
namely
the state $(1,1,0,0;0,0,-1,1)$. In this strong coupling limit this
state is not only massless at $T=U=1$ but
also on the following critical line
\beqa U=-i{1+iS_2T\over S_2-iT}.\label{newline}
\eeqa
This line contains the point $(T,U)=(1,1)$ for all possible values of $S_2$.
For $S_2=0$ this line becomes $T=1/U$. For $S_2=1$ one also obtains a
special line in the $T,U$ moduli space: consider, for example, $T_2={\rm Im}T=
0$. Then $U$ lies on the unit circle, ${\rm Re}U^2+{\rm Im}U^2=1$,
which is the boundary of the $U$ moduli space.
A similar discussion holds for all the other states in eq.(\ref{states}).
The associated critical line is
obtained from (\ref{newline}) by a corresponding $T/U$-duality
transformation.

Next consider a state listed in (\ref{statesr}), for example
$(1,1,0,0;1,0,1,-1)$.
In the strong coupling limit it is massless at the line
\beqa
U={1-S_2-iS_2T\over T-iS_2},\label{newliner}
\eeqa
which contains the point
$T=\bar U=\rho$. For $S_2=0$ it becomes $U=1/T$, and for $S_2=1$ this relation
is again satisfied for $U$ lying on the unit circle if $T$ lies on the
boundary of the moduli space, i.e. $T_2=1/2$.

\end{document}